\documentclass[aps,superscriptaddress,amssymb,showpacs,prb,twocolumn]{revtex4}
\usepackage{graphicx}
\begin{document}
%\myspace

\title{Optimization by Quantum Annealing: Lessons from Simple Cases}

\author{Lorenzo Stella}
\affiliation{International School for Advanced Studies (SISSA), and INFM
Democritos National Simulation Center, Via Beirut 2-4, I-34014 Trieste,
Italy}
\author{Giuseppe E. Santoro}
\affiliation{International School for Advanced Studies (SISSA), and INFM
Democritos National Simulation Center, Via Beirut 2-4, I-34014 Trieste,
Italy}
\affiliation{International Centre for Theoretical Physics
(ICTP), P.O.Box 586, I-34014 Trieste, Italy}
\author{Erio Tosatti}
\affiliation{International School for Advanced Studies (SISSA), and INFM
Democritos National Simulation Center, Via Beirut 2-4, I-34014 Trieste,
Italy}
\affiliation{International Centre for Theoretical Physics
(ICTP), P.O.Box 586, I-34014 Trieste, Italy}

\date{\today}

\begin{abstract}

This paper investigates the basic behavior and performance of simulated quantum 
annealing (QA) in comparison with classical annealing (CA). Three simple one 
dimensional case study systems are considered, namely a parabolic well, a double 
well, and a curved washboard. The time dependent Schr\"odinger evolution in either 
real or imaginary time describing QA is contrasted with the Fokker Planck evolution of CA. 
The asymptotic decrease of excess energy with annealing time is studied in each case,
and the reasons for differences are examined and discussed. The Huse-Fisher classical
power law of double well CA is replaced with a different power law in QA. The multi-well
washboard problem studied in CA by Shinomoto and Kabashima and leading classically
to a logarithmic annealing even in the absence of disorder, turns to a power law
behavior when annealed with QA. 
The crucial role of disorder and localization is briefly discussed.

\end{abstract}

\maketitle

%---------------------------------------------------------------------------
\section{Introduction}
%---------------------------------------------------------------------------

The idea of quantum annealing (QA) is a late offspring of the celebrated
{\em simulated thermal annealing} by Kirkpatrick {\it et al.} \cite{Kirk_CA}. 
In simulated annealing, the problem of minimizing a certain cost (or energy) function in a
large configuration space is tackled by the introduction of a fictitious
temperature, which is slowly lowered in the course of a Monte Carlo or 
Molecular Dynamics simulation \cite{Kirk_CA}.
This device allows an exploration of the configuration
space of the problem at hand, effectively avoiding trapping at unfavorable
local minima through thermal hopping above energy barriers. It makes for 
a very robust and effective minimization tool,
often much more effective than standard, gradient-based, minimization methods.

An elegant and fascinating alternative to such a classical simulated annealing (CA)
consists in helping the system escape the local minima through {\em quantum mechanics},  
by tunneling through the barriers rather than thermally overcoming them 
\cite{Finnila,QA_jpn}.
Experimental evidence in disordered Ising ferromagnets subject to transverse magnetic
fields showed that this strategy is not only feasible but presumably winning in
certain cases \cite{aeppli}.
In essence, in quantum annealing one supplements the classical energy function, let us
denote it by $H_{cl}$, with a suitable {\em time-dependent} quantum kinetic 
term, $H_{kin}(t)$, which is initially very large, for $t\le 0$, then gradually 
reduced to zero in a time $\tau$. 
The quantum state of the system $|\Psi(t)\rangle$, initially prepared in
the fully quantum ground state $|\Psi_0\rangle$ of $H(t=0)=H_{cl}+H_{kin}(0)$, 
evolves according to the Scr\"odinger equation 
\begin{equation} \label{Schrodinger:eqn}
i \hbar \frac{d}{dt}|\Psi(t)\rangle = [H_{cl}+H_{kin}(t)] |\Psi(t)\rangle \;,
\end{equation}
to reach a final state $|\Psi(t=\tau)\rangle$. 
A crucial basic question is then how the residual energy 
$\epsilon_{res}(\tau)=E_{fin}(\tau)-E_{opt}$, 
decreases for increasing $\tau$.
Here $E_{opt}$ is the absolute minimum of $H_{cl}$, and $E_{fin}(\tau)$ is the average
energy attained by the system after evolving for a time $\tau$, 
$E_{fin}(\tau)=\langle \Psi(\tau)| H_{cl} | \Psi(\tau)\rangle/
\langle \Psi(\tau)| \Psi(\tau)\rangle$, 
Generally speaking, this question has to do with the {\em adiabaticity} of the 
quantum evolution, i.e., whether the system is able, for sufficiently slow 
annealing (sufficiently long $\tau$), to follow the instantaneous ground state of 
$H(t)=H_{cl}+H_{kin}(t)$, for a judiciously chosen $H_{kin}(t)$.
The fictitious kinetic energy $H_{kin}(t)$ can be chosen quite freely, with 
the only requirement of being reasonably easy to implement.
For this reason, this approach has also been called Quantum Adiabatic 
Evolution \cite{Farhi_Science}.
 
At the level of practical implementations on an ordinary (classical) computer, 
the task of following the time-dependent Schr\"odinger evolution in Eq.\ \ref{Schrodinger:eqn}
is clearly feasible only for toy models with a sufficiently manageable Hilbert space
\cite{Farhi_Science}. 
Actual optimization problems of practical interest usually involve astronomically large Hilbert
spaces, a fact that calls for alternative stochastic (Quantum Monte Carlo) approaches.
These QMC techniques, in turn, are usually suitable to using {\em imaginary time}
quantum evolution, where the $i\hbar \partial_t$ in Eq.\ \ref{Schrodinger:eqn} is 
replaced by $-\hbar\partial_t$. One of the questions we will try to address in 
the present paper will be, therefore, if working with an imaginary-time Schr\"odinger 
evolution changes the quantum adiabatic evolution approach in any essential way.
We will argue that, as far as annealing is concerned, imaginary-time is 
essentially equivalent to real-time, and, as a matter of fact, can be quantitatively better.

Alternatively, a number of recent theoretical papers have applied
Path-Integral Monte Carlo strategies to QA. A certain success has been
obtained in several optimization problems, such as the folding of off-lattice 
polymer models \cite{Berne1,Berne2}, the random Ising model ground state 
problem \cite{science,prb_martonak}, and the Traveling Salesman Problem \cite{pre_martonak}. 
It is fair to say, however, that there is no general theory predicting the
performance of a QA algorithm, in particular correlating it with the 
energy landscape of the given optimization problem.
This is a quite unpleasant situation, in view of the fact that it is a-priori not obvious or
guaranteed that a QA approach should do better than, for instance, CA.
Indeed, for the interesting case of Boolean Satisfiability problems -- more precisely, 
a prototypical NP-complete problem  such as 3-SAT -- recent attempts in our group 
showed that Path-Integral Monte Carlo annealing may perform definitely {\em worse} 
than simple CA \cite{demian}. 

Evidently, the performance of QA over CA depends in detail on the energy landscape 
of the problem at hand, in particular on the nature and type of barriers separating 
the different local minima, a problem about which very little is known in many 
practical interesting cases \cite{TSP_landscape}.
It makes sense therefore to move one step back and concentrate attention 
on the simplest textbook problems where the energy landscape is well under control: 
essentially, one-dimensional potentials, starting from a double well potential,
the simplest form of barrier. On these well controlled landscapes we can carry 
out a detailed and exhaustive comparison between quantum adiabatic Schr\"odinger 
evolution, both in real and in imaginary time, and its classical deterministic 
counterpart, i.e., Fokker-Planck evolution. 

The rest of the paper is organized as follows. 
In Sec.\ \ref{Annealing_def:sec} we define the problem we want to tackle, i.e., comparing
Fokker-Planck annealing to Schr\"odinger annealing.
In Sec.\ \ref{tls:sec} we consider in detail the case of a double-well barrier. 
In Sec.\ \ref{washboard:sec} we move to the case of a potential with many minima, but
no disorder, were the behavior of classical and quantum annealing is remarkably
different. 
In Sec.\ \ref{disorder:sec} the crucial role of disorder is discussed. 
Finally, Sec.\ \ref{summary:sec} contains a summary of the results found
and a few concluding remarks. 
Details of the calculations are contained in the Appendices.

%------------------------------------------------------------------------
\section{Schr\"odinger versus Fokker-Planck annealing: Statement of the problem}
\label{Annealing_def:sec} 
%------------------------------------------------------------------------

Suppose we are given a potential $V(x)$, (with $x$ a Cartesian vector of 
arbitrary dimension), of which we need to determine the absolute minimum
($x_{opt}$,$E_{opt}=V(x_{opt})$). Assume generally a situation in which a 
steepest-descent approach, i.e., the strategy of following the gradient of $V$, would 
lead to trapping into one of the many local minima of $V$, and would thus not work.
Classically, as an obvious generalization of a steepest-descent approach, 
one could imagine of performing a stochastic (Markov) dynamics in x-space 
according to a Langevin equation 
\begin{equation} \label{Langevin}
\dot{x} = -\frac{1}{\eta(T)} \nabla V(x) + \xi(t) \;,
\end{equation}
where the strength of the noise term $\xi$ is controlled by the squared correlations
$\overline{\xi_i(t)\xi_j(t')}=2D(T)\delta_{ij}\delta(t-t')$, with $\bar{\xi}=0$.
Both $D(T)$ and $\eta(T)$  -- with dimensions of a diffusion constant and of a 
friction coefficient and related, respectively, to fluctuations and dissipation in the system -- 
are temperature dependent quantities which can be chosen, for the present optimization purpose,
with a certain freedom. The only obvious constraint is in fact that the correct 
thermodynamical averages should be recovered from the Langevin dynamics only if 
$\eta(T) D(T) = k_B T$, an equality known as Einstein's relation \cite{einstein:nota}.
Physically, $D(T)$ should be an increasing function of $T$, so as to lead
to increasing random forces as $T$ increases, with $D(T=0)=0$, since noise is 
turned off at $T=0$. 
Classical annealing can in principle be performed through this Langevin dynamics,
by slowly decreasing the temperature $T(t)$ as a function of time, from some 
initially large value $T_0$ down to zero. 
Instead of working with the Langevin equation -- a stochastic differential 
equation -- one might equivalently address the problem by studying 
the probability density $P(x,t)$ of finding a particle at position $x$ at time $t$. 
The probability density is well known to obey a {\em deterministic} time-evolution 
equation given by the {\em Fokker-Planck} (FP) equation \cite{vanKampen}:
\begin{equation} \label{FP:eqn}
\frac{\partial}{\partial t} P(x,t) 
\,=\, \frac{1}{\eta(T)} \, {\rm div} \left( P \nabla V \right) \,+\, 
D(T) \nabla^2 P \;.
\end{equation}
Here, the second term in the right-hand side represents the well known 
{\em diffusion term}, proportional to the diffusion coefficient $D(T)$, 
whereas the first term represents the effect of the {\em drift} force 
$-\nabla V$, inversely proportional to the friction coefficient
$\eta(T)=k_BT/D(T)$ \cite{einstein:nota}.
Annealing can now be performed, in principle, by keeping the system
for a long enough equilibration time at a large temperature $T_0$, 
and then gradually decreasing $T$ to zero as a function of time, $T(t)$, 
in a given annealing time $\tau$. 
We can model this by assuming
\begin{equation}
T(t) = T_0 f(t/\tau) \;, 
\end{equation}
where $f(y)$ is some assigned monotonically decreasing function for
$y\in [0,1]$, with $f(y\le 0)=1$ and $f(1)=0$. 
In this manner the diffusion constant $D$ in Eq.~(\ref{FP:eqn}) becomes 
a time-dependent quantity, $D_t=D(T(t))$. 
The FP equation should then be solved with an initial condition given by
the equilibrium Boltzmann distribution at temperature $T(t=0)=T_0$, i.e.,
$P(x,t=0)=e^{-V(x)/k_BT_0}$. 
The final average potential energy after annealing in excess of the true
minimum value, will then be simply given by:
\begin{equation} \label{residual:eqn}
\epsilon_{res}(\tau) = \int dx \, V(x) \, P(x,t=\tau) \,-\, E_{opt} \ge 0 \;,
\end{equation}
where $E_{opt}$ is the actual absolute minimum of the potential $V$.

In a completely analogous manner, we can conceive using Schr\"odinger's 
equation to perform a deterministic quantum annealing (QA) evolution of the 
system, by studying:
\begin{equation}
\xi \hbar \frac{\partial}{\partial t} \psi(x,t) = 
\left[ -\Gamma(t) \nabla^2 + V(x) \right] \psi(x,t) \;,
\end{equation}
where $\xi=i$ for a {\em real-time} (RT) evolution, while $\xi=-1$ for an
{\em imaginary-time} (IT) evolution. 
Here $\Gamma(t)=\hbar^2/2m_t$ will be our annealing parameter, playing the role
that the temperature $T(t)$ had in classical annealing.
Once again we may take $\Gamma(t)$ varying from some large value 
$\Gamma_0$ at $t\le 0$  -- corresponding to a small mass of the particle, hence to large 
quantum fluctuations -- down to $\Gamma(t=\tau)=0$, 
corresponding to a particle of infinite mass, hence without quantum 
fluctuations.
Again, we can model this with $\Gamma(t)=\Gamma_0 \, f(t/\tau)$, where $f$
is a preassigned monotonically decreasing function.
A convenient initial condition here will be $\psi(x,t=0)=\psi_0(x)$, 
where $\psi_0(x)$ is the {\em ground state} of the system at $t\le 0$, 
corresponding to the large value $\Gamma(t)=\Gamma_0$ and hence to
large quantum fluctuations. 
For such a large $\Gamma$ the ground state will be isolated, and separated by 
a large energy gap from all excited states.
The residual energy after annealing will be similarly given by 
Eq.~(\ref{residual:eqn}), where now, however, the probability $P(x,t=\tau)$
should be interpreted as 
\[ 
P(x,t)= \frac{|\psi(x,t)|^2}{\int dx^{\prime} \, |\psi(x^{\prime},t)|^2} \;.
\]
In general, the residual energy will be different for a RT or an IT 
Schr\"odinger evolution. We will discuss in some detail RT versus 
IT Schr\"odinger evolution in Sec.\ \ref{RT_vs_IT:sec}.

The basic question we pose is which annealing scheme is eventually more effective,
leading to the smallest final residual energies $\epsilon_{res}(\tau)$. 
This might seem an ill-posed question, because the time scales involved
in the classical and in the quantum evolution are different, and also
because, practically, the two approaches might involve different 
computational costs which would imply different CPU time scales.
In other words, it might seem that it only makes sense to ask
how well an annealing scheme performs in a given CPU-time $T_{CPU}$,
with all the unavoidable uncertainty associated to a CPU-time-related answer
(involving, among other things, the programmer's skills, the algorithmic
choices, and the computer architecture).
We will show, however, that the behavior of $\epsilon_{res}(\tau)$ can
be so vastly different for the different schemes, obviously in strict
relation with the form of the potential, that such time scale concerns
are often practically irrelevant \cite{timescale:nota}. 

%------------------------------------------------------------------------
\subsection{The harmonic potential: a warm-up exercise}
\label{harmonic:sec}
%------------------------------------------------------------------------

Preliminary to any further treatment of a potential with barriers, 
and as a warm-up exercise which will be useful later on,
we start here with the simple case of a parabolic potential in one 
dimension, $V(x)=kx^2/2$, which has a trivial minimum in $x=0$, 
with $E_{opt}=0$, and no barriers whatsoever. 

Let us consider classical FP annealing first. As detailed in the Appendix,
it is a matter of simple algebra to show that, for the harmonic potential, 
one can write a simple closed linear differential equation \cite{Shi_Kab} 
for the average potential energy $\epsilon_{pot}(t)$, which has the form:
\begin{equation} \label{der_residual_closed:eqn}
\frac{d}{dt} \epsilon_{pot}(t) = 
k D_t \left[ 1 - \frac{2}{k_B T(t)} \epsilon_{pot}(t) \right] \;, 
\end{equation}
the initial condition being simply given by the equipartition value
$\epsilon_{pot}(t=0)=k_BT_0/2$. 
As with every one-dimensional linear differential equation, 
Eq.~(\ref{der_residual_closed:eqn}) can be solved by quadrature for
any choice of $T(t)$ and $D_t=D(T(t))$. 
Assuming the annealing schedule to be parameterized by an exponent 
$\alpha_T>0$, $T(t)=T_0 (1-t/\tau)^{\alpha_T}$, $\tau$ being the annealing
time, and the diffusion constant $D(T)$ to behave as a power law of 
temperature, $D(T)=D_0 (T/T_0)^{\alpha_D}$ with $\alpha_D\ge 0$, we can 
easily extract from the analytical solution for $\epsilon_{pot}(t)$ 
the large $\tau$ asymptotic behavior of the final residual energy
$\epsilon_{res}(\tau)=\epsilon_{pot}(t=\tau)$.  
That turns out to be:  
\begin{equation} \label{CA_expon_2:eqn}
\epsilon_{res}(\tau) \approx \tau^{-\Omega_{CA}} \hspace{4mm} \mbox{with}
\hspace{4mm}
\Omega_{CA} = \frac{\alpha_T}{\alpha_T(\alpha_D-1)+1} \;.
\end{equation}
Trivial as it is, annealing proceeds here extremely fast, with a power-law 
exponent $\Omega_{CA}$ that can increase without bounds (for instance 
if $\alpha_D=1$) upon increasing the exponent $\alpha_T$ of the annealing
schedule.

Consider now the Schr\"odinger evolution problem for this potential, 
\begin{eqnarray} \label{Sch_R:eqn}
&& \xi \frac{\partial}{\partial t} \psi(x,t) = 
\left[ -\Gamma(t) \nabla^2 + \frac{k}{2}x^2 \right] \psi(x,t) \\
&& \psi(x,t=0) = \psi_{0}(x) \;, \nonumber 
\end{eqnarray}
where $\psi_{0}(x) \propto \exp{(-B_0 x^2/2)}$ is the ground state Gaussian 
wavefunction corresponding to the initial value of the Laplacian coefficient 
$\Gamma(t=0)=\Gamma_0$, 
and $\xi=i\hbar$ or $\xi=-\hbar$ for a real time (RT) or an imaginary time 
(IT) evolution, respectively.
This problem is studied in detail in the Appendix, where we show that
a Gaussian {\it Ansatz} for $\psi(x,t)$, of the form
$\psi(x,t) \propto \exp{(-B_t x^2/2)}$ with $\mbox{Real}(B_t)>0$, satisfies
the time-dependent Schr\"odinger equation as long as the inverse variance 
$B_t$ of the Gaussian satisfies the following ordinary non-linear first-order 
differential equation:
\begin{eqnarray} \label{Gauss_ODE_B:eqn}
-\xi \dot{B_t} &=& k - 2 \Gamma(t) B_t^2  \\
B_{t=0} &=& B_0 = \sqrt{ \frac{k}{2\Gamma_0} } \;. \nonumber
\end{eqnarray}
Contrary to the classical case, there is no simple way of recasting the 
annealing problem in terms of a closed linear differential equation for the
average potential energy $\epsilon_{pot}(t)$. 
The final residual energy $\epsilon_{res}(\tau)=\epsilon_{pot}(t=\tau)$ is
still expressed in terms of $B_{\tau}$ 
(or better, of its real part $\Re(B_{\tau})$), 
\begin{equation}
\epsilon_{res}(\tau) = \frac{\int dx \, V(x) |\psi(x,t=\tau)|^2}
                            {\int dx \, |\psi(x,t=\tau)|^2} =
\frac{k}{4\Re(B_{\tau})} \;,
\end{equation}
but the behavior of $B_t$ must be extracted from the study of the 
non-linear equation (\ref{Gauss_ODE_B:eqn}). 
The properties of the solutions of Eq.~(\ref{Gauss_ODE_B:eqn}) are studied 
in detail in the Appendix, where we show that: 
\begin{itemize}
\item[i)] 
$\epsilon_{res}(\tau)$ cannot decrease faster than $1/\tau$, for
large $\tau$, i.e., a power-law exponent
$\epsilon_{res}(\tau) \approx \tau^{-\Omega_{QA}}$ is bound to be
$\Omega_{QA}\le 1$.
\item[ii)] 
Adopting a power-law annealing schedule 
$\Gamma(t)=\Gamma_0(1-t/\tau)^{\alpha_{\Gamma}}$, the exponent 
$\Omega_{QA}$ for the IT case is 
\begin{equation} \label{QA_expon:eqn}
\Omega_{QA} = \frac{\alpha_{\Gamma}}{\alpha_{\Gamma}+2} \;,
\end{equation}
increasing towards the upper bound $1$ as $\alpha_{\Gamma}$ is increased
towards $\infty$.
\item[iii)]
RT quantum annealing proceeds with exactly the same exponent $\Omega_{QA}$ as IT 
quantum annealing -- although $\epsilon_{res}^{RT}(\tau) \ge \epsilon_{res}^{IT}(\tau)$ in
general --, except that the limit $\alpha_{\Gamma}\to \infty$ (abrupt 
switch-off of the Laplacian coefficient) is singular in the RT case.
%
%(NOTA: Indeed, if we take $\Gamma(t>0)=0$, the RT evolution performs no annealing 
%at all, since the real part of $B_t$ is constant in time, so that 
%$\Omega_{QA}^{RT}=0$ and not $1$, as instead happens in the limit of 
%very large $\alpha_{\Gamma}$.)
%
\end{itemize}

Summarizing, we have learned that, for a single parabolic valley in configuration space,
both CA and QA proceed with power-laws, but CA can be much more efficient 
than QA, with an arbitrarily larger power-law exponent. 
We underline however that this is merely an academic matter at this point,
steepest descent being much more efficient than both CA and QA in such 
a simple case. 
The power of QA shows up only when potentials with barriers are considered.

%------------------------------------------------------------------------
\section{The simplest barrier: a double-well potential} 
\label{tls:sec}
%------------------------------------------------------------------------

Let us take now the classical potential to be optimized as a simple
double-well potential in one-dimension
\begin{equation} \label{Vsym:eqn}
V_{\rm sym}(x) = V_0 \frac{ (x^2-a^2)^2 }{a^4} + \delta x \;,
\end{equation}
with $V_0$, $a$, and $\delta$ real constants.
In absence of the linear term ($\delta=0$), the potential has two 
degenerate minima located at $\pm a$, and separated by a barrier of height $V_0$.
When a small linear term $\delta>0$ is introduced , with $\delta a << V_0$, 
the two degenerate minima are split by a quantity $\Delta_V \approx 2\delta a$, 
the minimum at $x\approx -a$ becoming slightly favored. 
For reasons that will be clear in a moment, it is useful to slightly
generalize the previous potential to a less symmetric situation,
where the two wells possess definitely distinct {\em curvatures} at
the minimum (i.e, their widths differ substantially). 
This is realized easily, with a potential of the form:
\begin{equation} \label{Vasym:eqn}
V_{\rm asym}(x) = 
\left\{ \begin{array}{c} 
         V_0 \frac{(x^2-a_+^2)^2}{a_+^4} + \delta x 
             \hspace{10mm} \mbox{for} \; x\ge0 \\
         V_0 \frac{(x^2-a_-^2)^2}{a_-^4} + \delta x 
             \hspace{10mm} \mbox{for} \; x < 0 \\
       \end{array}
\right. \;,
\end{equation}
with $a_+\ne a_-$, both positive. 
(The discontinuity in the second derivative at the origin is of no consequence 
in our discussion.)
To linear order in the small parameter $\delta$, the two minima are
now located at $x_{\pm}=\pm a_{\pm} - \delta a_{\pm}^2/(8V_0)$, 
the splitting between the two minima is given by 
$\Delta_V = \delta (a_{+}+a_{-})$, while the second derivative of the 
potential at the two minima, to lowest order in $\delta$, is given by:
\[ 
V^{\prime\prime}(x=x_{\pm}) = \frac{8V_0}{a_{\pm}^2} \;. 
\]
Obviously, $V_{\rm sym}$ is recovered if we set $a_{+}=a_{-}=a$ in 
$V_{\rm asym}$.

We now present the results obtained by the annealing schemes introduced
in Sec.~\ref{Annealing_def:sec} above. 
The Fokker-Planck and the Schr\"odinger equation (both in RT and in IT) were 
integrated numerically using a fourth-order Runge-Kutta method, 
after discretizing the $x$ variable in a sufficiently fine real space grid \cite{grid}. 
For the FP classical annealing, the results shown are obtained with a linear temperature 
schedule, $T(t)=T_0(1-t/\tau)$ (i.e., $\alpha_T=1$), and a diffusion coefficient simply 
proportional to $T(t)$, $D_t=D_0(1-t/\tau)$ (i.e., $\alpha_D=1$). 
Consequently, the friction coefficient is kept constant in $t$, 
$\eta_t=k_B T(t)/D_t = k_B T_0/D_0$. 
Similarly, for the Schr\"odinger quantum annealing we show results obtained with
a coefficient of the Laplacian $\Gamma(t)$ vanishing linearly in a time 
$\tau$, $\Gamma(t)=\Gamma_0(1-t/\tau)$ (i.e., $\alpha_{\Gamma}=1$). 
%
%%Dimensionless quantities have been defined accordingly (SPECIFY BETTER?).
%
%------------------------------------------------------------------------
\begin{figure}[!tbp]
  \includegraphics*[width=8cm,angle=0]{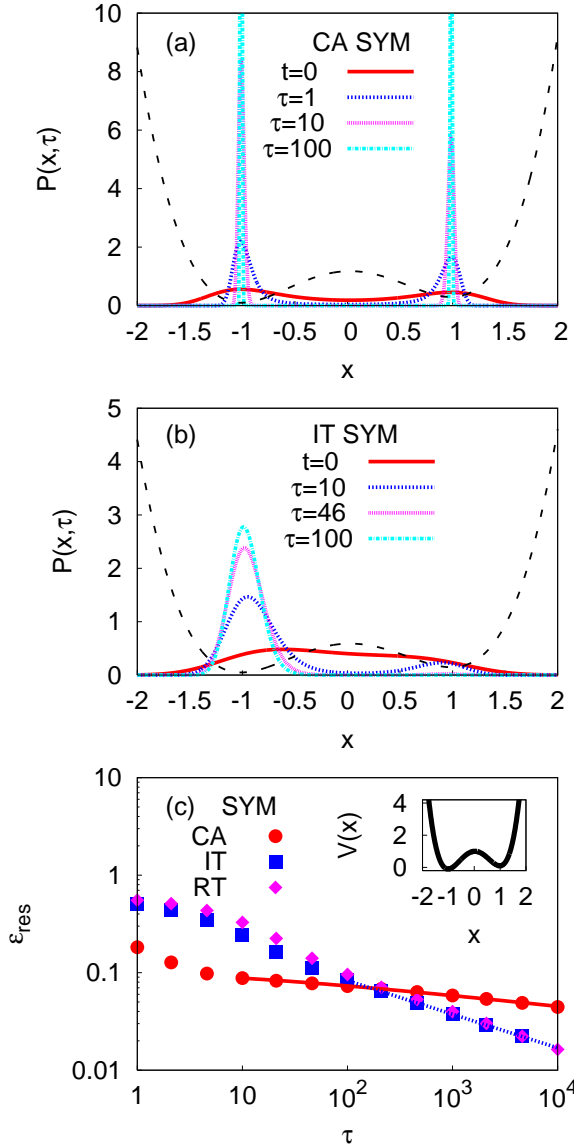}
  \caption{\label{fig_ann_sym:fig} 
   (a,b): The annealed final probability distribution $P(x,t=\tau)$
   at different values of the annealing time $\tau$, for both the
   Fokker-Planck classical annealing (CA, panel (a)), and the Imaginary Time
   Schr\"odinger quantum annealing (IT-QA, panel (b)). 
   (c) Final residual energy $\epsilon_{res}(\tau)$ versus annealing time
   $\tau$ for quantum annealing in Real Time (RT) and Imaginary Time (IT) 
   compared to the Fokker-Planck classical annealing (CA). 
   The solid line in (c) is a fit of the CA data.
   The double well potential (dashed line in (a,b), inset of (c)) 
   is here given by Eq.~(\protect\ref{Vasym:eqn}) with $a_{+}=a_{-}=a=1$. 
   }
\end{figure}
%------------------------------------------------------------------------
%------------------------------------------------------------------------
\begin{figure}[!tbp]
  \includegraphics*[width=8cm,angle=0]{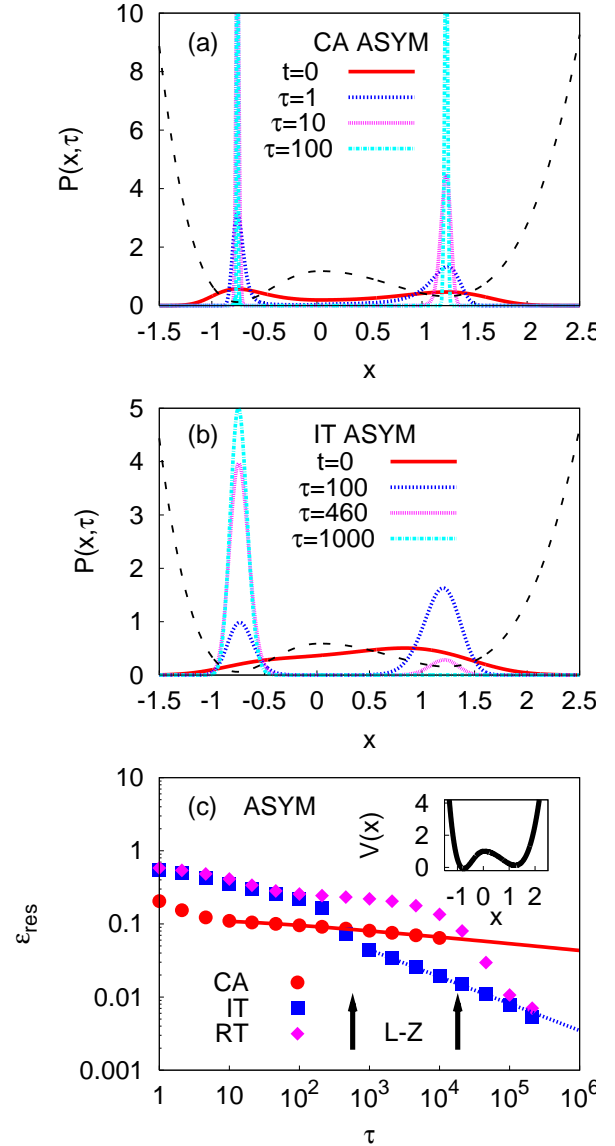}
  \caption{\label{fig_ann_asym:fig} 
   Same as Fig.\ \protect{\ref{fig_ann_sym:fig}}, for the asymmetric 
   potential in Eq.~(\protect\ref{Vasym:eqn}) with 
   $a_{+}=1.25, a_{-}=0.75$ (dashed line in (a,b), inset of (c)).
   Notice the different behavior of RT and IT, in the present case.
   }
\end{figure}
%------------------------------------------------------------------------
%------------------------------------------------------------------------
\begin{figure}[!tbp]
  \includegraphics*[width=8cm,angle=0]{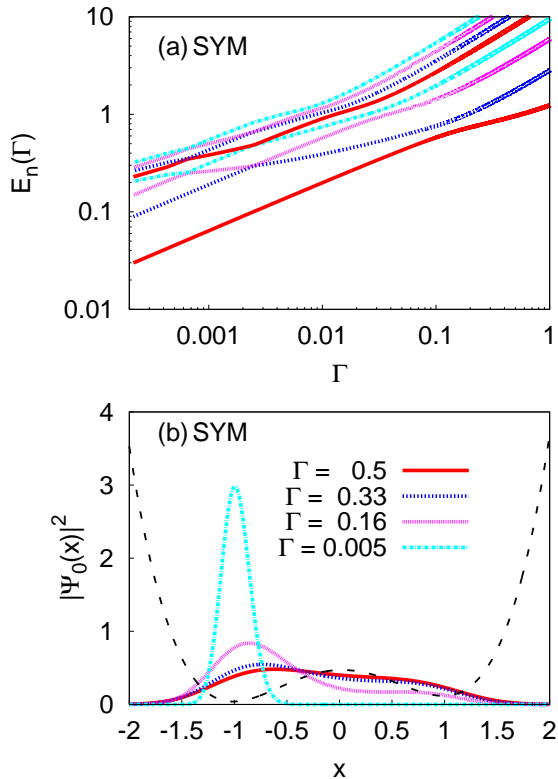}
  \caption{\label{fig_eigen_sym:fig} 
   Instantaneous eigenvalues (a) and ground state wavefunctions (b) 
   of the Schr\"odinger problem $H\psi=E\psi$ for different values of $\Gamma$,
   for the symmetric potential in Eq.~(\protect\ref{Vasym:eqn}) with 
   $a_{+}=a_{-}=a=1$.
   }
\end{figure}
%------------------------------------------------------------------------
%------------------------------------------------------------------------
\begin{figure}[!tbp]
  \includegraphics*[width=8cm,angle=0]{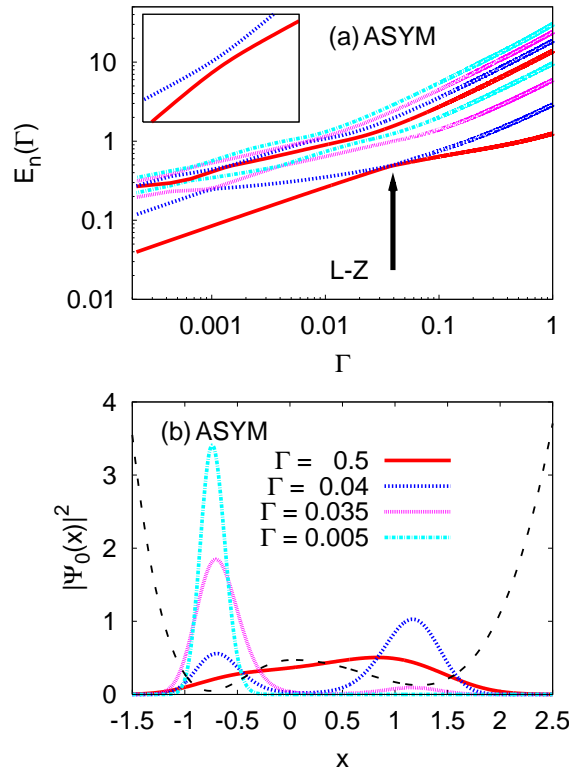}
  \caption{\label{fig_eigen_asym:fig} 
   Same as Fig.\ \protect{\ref{fig_eigen_sym:fig}}, for the asymmetric
   potential in Eq.~(\protect\ref{Vasym:eqn}) with 
   $a_{+}=1.25, a_{-}=0.75$. 
   Notice the clear Landau-Zener avoided crossing in (a), indicated by the
   arrow and magnified in the inset.
   }
\end{figure}
%------------------------------------------------------------------------
%%------------------------------------------------------------------------
%\begin{figure}[!tbp]
%  \includegraphics*[width=8cm,angle=0]{Figures/new2_fig_ann_6.eps}
%  \caption{\label{fig_overlap_sym:fig} 
%   Overlap between the imaginary time time-dependent state $|\Psi(t)\rangle$ 
%   and the first two instantaneous eigenstates at the corresponding value of
%   $\Gamma(t)$, showing that adiabaticity is always fulfilled except in the
%   very last stage of annealing.
%   }
%\end{figure}
%%------------------------------------------------------------------------
%%------------------------------------------------------------------------
%\begin{figure}[!tbp]
%  \includegraphics*[width=8cm,angle=0]{Figures/new2_fig_ann_5.eps}
%  \caption{\label{fig_overlap_asym:fig} 
%   Overlap between the imaginary time time-dependent state $|\Psi(t)\rangle$ 
%   and the first two instantaneous eigenstates at the corresponding value of
%   $\Gamma(t)$,  ...
%   }
%\end{figure}
%%------------------------------------------------------------------------
 
Fig.~\ref{fig_ann_sym:fig} shows the results obtained for the final
annealed probability distribution $P(x,t=\tau)$ at different values of $\tau$,
for both the Fokker-Planck (CA, panel (a)) and the Scr\"odinger imaginary-time case 
(IT, panel (b)), for a symmetric double well potential $V_{\rm sym}(x)$, 
with $V_0=1$ (our unit of energy), $a=a_+=a_-=1$ (unit of length), $\delta=0.1$. 
Fig.~\ref{fig_ann_sym:fig}(c) summarizes the results obtained for the residual 
energy $\epsilon_{res}(\tau)$. 
Fig.~\ref{fig_ann_asym:fig}(a,b,c) shows the corresponding results for an
asymmetric double well potential, Eq.~ \ref{Vasym:eqn}, with $a_+=1.25$, 
$a_-=0.75$, and $\delta=0.1$.
 
We notice immediately that QA wins, in both cases, over CA for large enough value 
of $\tau$. The RT-QA behaves as its IT counterpart for the symmetric double 
well, while it shows a different behavior in the asymmetric case (see below for comments).
To go deeper into the details of the different evolutions, let us begin discussing the 
CA data (panel (a) and (c) of Figs.~\ref{fig_ann_sym:fig} and \ref{fig_ann_asym:fig}), 
which show similar behaviors for both choices of the potential.
Starting from an initially broad Boltzmann distribution at a high $T=T_0=V_0$, 
$P(x,t=0)$ (solid lines), the system quickly sharpens the distribution 
$P(x,t)$ into two well-defined and quite narrow peaks located around the 
two minima $x_{\pm}$ of the potential. This agrees very well with
expectations based on the CA in a harmonic potential, which showed that
the width of the Gaussian should decrease linearly in $\tau$
($\Omega_{CA}=1$ for $\alpha_T=\alpha_D=1$), as is indeed found in our
double well case too.
If we denote by $p_{\pm}$ the integral of each of the two narrow peaks, with
$p_{-}+p_{+}=1$, it is clear that the problem has effectively
been reduced to a {\em discrete} two-level system problem. 
The time evolution $p_{\pm}$ therefore obeys a discrete Master equation which involves
the thermal promotion of particles over the barrier $V_0$, of the form:
\begin{equation} \label{HF_1:eqn}
\gamma^{-1} \frac{dp_{+}}{dt} = [1-p_{+}(t)] \, e^{-\frac{\Delta_V+B}{k_BT(t)}} -
p_{+}(t) e^{-\frac{B}{k_BT(t)}}  \;,
\end{equation}
where $\gamma$ is an attempt frequency, 
while $B=V_0-V(x_{+})$ and $B+\Delta_V=V_0-V(x_{-})$ are the potential barriers 
seen from a particle in the metastable minimum, $x_{+}$, and in the true 
minimum, $x_{-}$, respectively.
Eq.~(\ref{HF_1:eqn}) was studied by Huse and Fisher in 
Ref.~\onlinecite{Huse_Fisher}, where they showed that the asymptotic value
of the residual energy $\epsilon_{res}(\tau)=\Delta p_{+}(\tau)$ is given by:
\begin{equation} \label{HF_2:eqn}
\epsilon_{res}(\tau) % = \Delta_V p_{+}(\tau) 
\sim {\rm const} \; \left( \gamma\tau \right)^{- \frac{\Delta_V}{B} }
\left( \ln{\tilde{\gamma}\tau} \right)^{2 \frac{\Delta_V}{B} } \;,
\end{equation}
where $\tilde{\gamma}$ is a constant.
%
%%
%%\sim \Delta \Gamma(1+\Delta/B)\left( \frac{T_0}{B\gamma\tau} \right)^{\Delta/B}
%%\left( \frac{B}{\Delta} \ln{ \frac{T_0}{B\gamma\tau} } \right)^{2\Delta/B}
%%-\Delta \left( \frac{T_0}{B\gamma\tau} \right)^{2\Delta/B}
%%\frac{1}{1+(T_0/B\gamma\tau)^{\Delta/B}} \;.
%
So, apart from logarithmic corrections, the leading power-law behavior
is of the form $\epsilon_{res}\sim \tau^{-\Delta_V/B}$, where the exponent 
is controlled by the ratio $\Delta_V/B$ between the energy splitting of the
two minima $\Delta_V$ and the barrier $B$. 
As shown in Figs.\ \ref{fig_ann_sym:fig}(c) and \ref{fig_ann_asym:fig}(c) 
(solid lines through solid circles), 
the asymptotic behavior anticipated by Eq.~(\ref{HF_2:eqn}) fits nicely our 
CA residual energy data (solid circles), as long as the logarithmic corrections are 
accounted for in the fitting procedure \cite{fit}.
Obviously, we can make the exponent as small as we wish by reducing 
the linear term coefficient $\delta$, and hence the ratio $\Delta_V/B$, 
leading to an exceedingly slow classical annealing.

The behavior of the QA evolution is remarkably different. 
Starting from Fig.~\ref{fig_ann_sym:fig}, we notice that IT and RT evolutions give
very similar residual energies, definitely faster decaying than the CA data, while
the corresponding final wavefunction only slowly narrows around the minimum of the
potential. 
Notice also the asymptotic behavior of the residual energy, $\epsilon_{res}(\tau)\propto 
\tau^{-1/3}$, indicated by the dashed line in Fig.~\ref{fig_ann_sym:fig}(c): this rather
strange exponent is simply the appropriate one for the Schr\"odinger annealing with
a linear schedule within an harmonic potential 
(the lower minimum valley, see Sec.~\ref{harmonic:sec}).
The asymmetric potential results, shown in Fig.~\ref{fig_ann_asym:fig}, 
are even more instructive. 
The initial wavefunction squared $|\psi(x,t=0)|^2$ corresponds to a
quite small mass (a large $\Gamma_0=0.5$), and is broad and delocalized over
both minima (solid line).
As we start annealing, and if the annealing time $\tau$ is relatively short 
-- that is, if $\tau<\tau_{c}$, with a characteristic time $\tau_{c}$ which depends on
which kind of annealing, RT or IT, we perform -- the final wavefunction becomes 
mostly concentrated on the {\em wrong minimum}, roughly corresponding to the 
ground state with a still relatively large $\Gamma_1<\Gamma_0$ 
(see Fig.~\ref{fig_eigen_asym:fig}). 
The larger width of the wrong valley is crucial, giving a smaller quantum kinetic energy 
contribution, so that tunneling to the other (deeper) minimum does not yet occur. 
By increasing $\tau$, there is a crossover: the system finally recognizes 
the presence of the other minimum, and effectively tunnels into it, with a
residual energy that, once again, decays asymptotically as 
$\epsilon_{res}(\tau)\propto \tau^{-1/3}$ (dashed line in Fig.~\ref{fig_ann_asym:fig}(c)).
There is a characteristic annealing time $\tau_{c}$ -- different in the two Scr\"odinger
cases, RT and IT -- above which tunneling occurs, 
and this shows up as the clear crossover in the residual energy behavior of both IT and
RT, shown in Fig.\ \ref{fig_ann_asym:fig}(c).

These findings can be quite easily rationalized by looking at the {\em instantaneous} 
(adiabatic) eigenvalues and eigenstates of the associated time-independent Schr\"odinger 
problem, which we show in Fig.\ \ref{fig_eigen_asym:fig}(a,b). 
Looking at the instantaneous eigenvalues shown in Fig.\ \ref{fig_eigen_asym:fig}(a) 
we note a clear avoided-crossing occurring at $\Gamma=\Gamma_{LZ}\approx 0.038$,
corresponding to a resonance condition between the states in the two different
valleys of the potential. For $\Gamma>\Gamma_{LZ}$ the ground state wavefunction
is predominantly concentrated in the wider but metastable valley, while for 
$\Gamma<\Gamma_{LZ}$ it is mostly concentrated on the deeper and narrower global 
minimum valley.
In the full time-dependent RT evolution, transfer to the lower valley is a Landau-Zener
problem \cite{Landau,Zener}:
%%(the whole QA problem is just a cascade of them) \cite{science,prb_martonak}.
the characteristic time $\tau_{c}$ for the tunneling event is given by 
$\tau_{LZ}=\hbar\alpha\Gamma_0/2\pi\Delta^2$, where $\alpha$ is the relative slope of
of the two crossing branches as a function of $\Gamma$, $2\Delta$ is the gap at the
avoided-crossing point, and $\Gamma_0$ is the initial value of the annealing parameter.
(For the case shown in Fig.\ \ref{fig_eigen_asym:fig}, we have $2\Delta=0.0062$, 
$\alpha=2.3$, hence $\tau_{LZ}\approx 18980$, see rightmost arrow in 
Fig.~\ref{fig_ann_asym:fig}(c).)
The Landau-Zener probability of jumping, during the evolution, from the ground state 
onto the ``wrong'' (excited) state upon fast approaching of the avoided level crossing 
is $P_{ex}=e^{-\tau/\tau_{LZ}}$, so that adiabaticity applies only if the annealing 
is slow enough, $\tau>\tau_{LZ}$.
The IT characteristic time is smaller, in the present case, than the RT one.
We will comment further on this point in Sec.\ \ref{RT_vs_IT:sec}. 
In a nutshell, the reason for this is the following.
After the system has jumped into the excited state, which occurs with a probability 
$P_{ex}=e^{-\tau/\tau_{LZ}}$, the residual IT evolution will filter out the excited state; 
this relaxation towards the ground state is controlled by the annealing rate
as well as by the average gap seen during the residual evolution. 
Numerically, the characteristic time $\tau_c$ seen during the IT evolution is of the order of
$\hbar/(2\Delta)$, see leftmost arrow in Fig.~\ref{fig_ann_asym:fig}(c), rather than being 
proportional to $1/\Delta^2$ as $\tau_{LZ}$ would imply.

%
%% I would cut this part
%
%%In order to qualify the level of adiabaticity, it is instructive to follow, 
%%during the annealing evolution, the overlap between the actual 
%%time-evolved state at time $t$, $|\Psi(t)\rangle$, and the first few instantaneous 
%%eigenstates at the corresponding value of $\Gamma(t)$, $|\Psi_n^{\Gamma(t)}\rangle$,
%%as shown in Fig.\ \ref{fig_eigen_asym:fig}(c) for two different values of $\tau$. 
%%%
%%For fast annealing, $\tau=100<\tau_{LZ}$, the system follows for a while the
%%instantaneous ground state, but then ($t>0.92\tau$) it jumps onto the first excited
%%state. Eventually, it disperses onto a whole shower of higher excited states in the 
%%final part of the annealing ($t\approx 0.999\tau$).
%%For slow annealing, $\tau=1000>\tau_{LZ}$, on the contrary, the system still jumps
%%onto the first excited state at $t\approx0.92\tau$, but subsequently recovers 
%%a large overlap with the ground state at $t\approx 0.97\tau$, signalling that 
%%tunneling to the other minimum has effectively had time to occur.
%%We notice, in passing, that the symmetric potential results are similar in spirit,
%%but the presence of a Landau-Zener avoided crossing is not so sharp:
%%in a sense, the ground state is always slightly more concentrated on the actual minimum
%%at $x_{-}$, and it is separated by a sizeable gap from all excited states.
%

\begin{figure}[!tbp]
  \includegraphics*[width=8cm,angle=0]{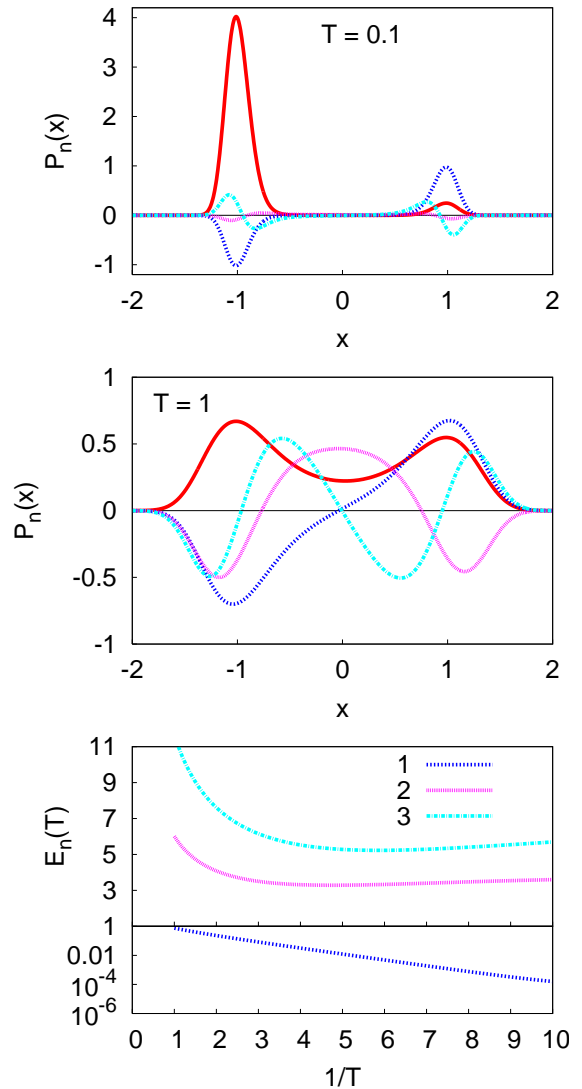}
  \caption{\label{fig_diag_SYM:fig} 
   Instantaneous Fokker-Planck eigenvalues (panel (c)) as a function of 
   temperature $T$, and the corresponding eigenstates for two values of $T$
   (panels (a) and (b)).
   The potential is here the symmetric one, $V_{\rm sym}$ in Eq.~\protect\ref{Vsym:eqn}  
   with $V_0=1$, $a=1$, $\delta=0.1$. 
   Similar results (not shown) are obtained for the asymmetric
   double well potential $V_{\rm asym}$. 
   }
\end{figure}

Obviously, instantaneous eigenvalues/eigenvectors can be studied for the Fokker-Planck 
equation as well; their properties, however, are remarkably different from 
the Landau-Zener scenario just described for the Schr\"odinger case. 
Fig.\ \ref{fig_diag_SYM:fig}(c) shows the first four low-lying
eigenvalues of the FP equation as a function of $T$ (for a symmetric choice of the potential),
while Fig.\ \ref{fig_diag_SYM:fig}(a,b) show the corresponding eigenstates for two
value of the temperature, $T/V_0=1$ and $T/V_0=0.1$. 
(The asymmetric potential cases are virtually identical, and are not shown). 
The lowest eigenvalue of the FP operator is identically $0$ and the corresponding
eigenvector \cite{vanKampen} is the Boltzmann distribution $e^{-V(x)/k_BT}$, 
with roughly symmetric maxima on the two valleys.
The first excited state correspond to distribution peaked on the two valley but with a node
at the origin, and is separated from the ground state by an exponential small
Arrhenius-like gap $e^{-B/k_BT}$. 
Higher excited states are separated by a very large gap, so that, effectively, only the
two lowest lying states dominate the dynamics at small temperature. 
The reduction of a continuum double-well FP classical dynamics onto
a discrete effectively quantum two-level system, previously noticed, is quite 
evident from this form of the spectrum. 
On the contrary, the true quantum case never allowed for a discrete two-level 
system description whatsoever, except perhaps for large $\Gamma$.
For small enough $\Gamma<\Gamma_{LZ}$, in particular, the tower of oscillator 
states within the valley at $x_{-}$ is always very close in energy to the actual 
ground state, and the quantum annealing evolution reduces effectively to a 
particle in a single harmonic well. This explains the rather large 
width of the final distributions $P(x,\tau)$ observed in the quantum case.
%
%------------------------------------------------------------------------
%------------------------------------------------------------------------

Summarizing, we have found that QA and CA proceed in a remarkably different way.
CA is sensitive to the height of the barrier, more precisely to the ratio $\Delta_V/B$
between the energy offset $\Delta_V$ of the two minima, and the barrier height $B$. 
On the contrary, QA crucially depends on the tunneling probability between the two 
valleys, which is reflected in a Landau-Zener (avoided crossing) gap: 
a wide tunneling barrier is obviously bad for QA. 
Finally, we noticed that RT and IT proceed with somewhat different characteristic times: 
we discuss this issue a bit more in the following section.
%
%If a given annealing time $\tau$ is available, the winner between QA and CA
%depends on the Landau-Zener scale $\tau_{LZ}$: if $\tau_{LZ}<\tau$, then tunnelling 
%will occur during the QA evolution, and QA will be faster, while for $\tau_{LZ}>\tau$ 
%CA will win out.
%%@ QUALCOSA NON MI TORNA QUI: PERCHE' NON ENTRANO I PARAMETRI DEL CA NELLA GARA?
%%  CI DEVE ESSERE UN'ALTRA CONDIZIONE AGGIUNTIVA, O QUALCOSA!
%% Vero ... Omettiamo ogni commento, abbiamo gia' detto cosa governa le varie
%% quantita'

%------------------------------------------------------------------------
\subsection{Real- versus imaginary-time Schr\"odinger evolution}
\label{RT_vs_IT:sec}
%------------------------------------------------------------------------
 
A Schr\"odinger dynamics in imaginary-time (IT)
is clearly much more convenient than that in real-time (RT) for simulations on 
current classical computers, but it makes a difference in the final results?
The answer to this question is, we believe: no, it does not make a difference, in
essence, although IT does give quantitatively better results. 

To qualify this statement, let us denote by $|\Psi^{(\xi)}(t)\rangle$ 
the solution of the Scr\"odinger equation
\begin{eqnarray}
\xi \frac{d}{dt}|\Psi^{(\xi)}(t)\rangle &=& 
[H_{cl}+H_{kin}(t)] |\Psi^{(\xi)}(t)\rangle \nonumber \\
|\Psi^{(\xi)}(t_0)\rangle &=& |\Psi_0\rangle  
\end{eqnarray}
where we assume that $|\Psi_0\rangle$ is the ground state of the initial
Hamiltonian at time $t_0$, $H_{cl}+H_{kin}(t_0)$, 
while $\xi=i\hbar$ for RT or $\xi=-1\hbar$ for IT. 
By definition, the final residual energy after annealing up
to time $t_f=t_0+\tau$, where the kinetic energy is finally turned off, is given by:
\begin{equation}
\epsilon^{(\xi)}_{res}(\tau)=
\frac{ \langle \Psi^{(\xi)}(t_0+\tau)| H_{cl} | \Psi^{(\xi)}(t_0+\tau)\rangle}
 { \langle \Psi^{(\xi)}(t_0+\tau)| \Psi^{(\xi)}(t_0+\tau)\rangle}-E_{opt} \;.
\end{equation}
We conjecture that the residual energies for the two alternative
way of doing a Schr\"odinger evolution verify the following: 
i) the IT residual energy is {\em not larger} then the RT one, that is
\begin{equation}
\epsilon^{(IT)}_{res}(\tau) \le \epsilon^{(RT)}_{res}(\tau) \;,
\end{equation}
and 
ii) in many problems, the leading asymptotic behavior, for $\tau\to \infty$, might be
identical for $\epsilon^{(IT)}_{res}(\tau)$ and $\epsilon^{(RT)}_{res}(\tau)$.

Expectation (i) seems very reasonable: it is simply inspired by the 
time-independent case, where it is well known that the IT Schr\"odinger 
dynamics tends to ``filter the ground state'' out of the initial trial 
wave function, as long as the gap between the GS and the first excited 
state is non-zero.
However, we have here a time-dependent situation, and the result is
a priori not guaranteed. We do not have a proof of this statement,
but we have verified it in all the cases where an explicit integration of the
Schr\"odinger equation has been possible (see, for instance, the results of the previous
Section). 
(Needless to say, we have no proof of (ii) either, but, again, it never failed in all 
our tests.)
 
\begin{figure}
  \includegraphics*[width=8cm,angle=0]{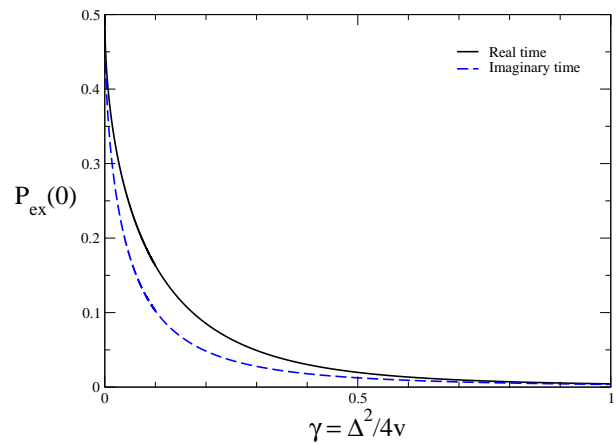}
  \caption{\label{zener:fig}
   The probability $P_{ex}(0)$ of ending up into the excited state, given by
   Eq.\ \protect{\ref{lz_2:eqn}}, for the discrete two-level system 
   problem in Eq.~ \protect{\ref{lz_1:eqn}}, for both imaginary-time
   (IT, dashed line) and real-time (RT, solid line) Schr\"odinger annealing.
   The large-$\gamma$ behavior of $P_{ex}(0)$ is, in both cases, given by
   $P_{ex}(0)\approx 1/(256\gamma^2)$. 
          }
\end{figure}

The simplest time-dependent problem where one can test our conjectures, is the discrete 
two-level system (TLS) problem. 
Here, in terms of Pauli matrices, $H_{cl}=\Delta \sigma^z$, while 
$H_{kin}(t)=-\Gamma(t) \sigma^x$, with $\Gamma(t)=-vt$. 
The full $H(t)$ is therefore
\begin{equation} \label{lz_1:eqn}
H(t) = \Delta \sigma^z - \Gamma(t) \sigma^x \;.
\end{equation}
The annealing interpretation is very simple:
the classical optimal state is $|\downarrow\rangle$, with energy 
$E_{opt}=-\Delta$, separated from the excited state $|\uparrow\rangle$ by 
a gap $2\Delta$. 
The kinetic term induces transitions between the two classical states.
Starting from the ground state of $H(t_0)$ at time $t_0=-\tau$ we let
the system evolve up to to time $t_f=t_0+\tau=0$, when the Hamiltonian is entirely classical, 
$H(t_f=0)=H_{cl}=\Delta \sigma^z$. 
The probability of missing the instantaneous final ground state 
$|\downarrow\rangle$, ending up with the excited state $|\uparrow\rangle$, is:
$P_{ex}(0)= |\langle \uparrow |\Psi^{(\xi)}(0)\rangle|^2/
         \langle \Psi^{(\xi)}(0)| \Psi^{(\xi)}(0)\rangle$. 
In principle, $P_{ex}$ depends, for given $\Delta$, both on the initial 
$\Gamma(t_0)=\Gamma_0=v\tau$ and on the annealing time $\tau$. 
The really important parameter, however, turns out to be the ratio $v$ between these
two quantities, which determines the ``velocity of annealing'': 
Taking $\tau\to \infty$ (i.e., $t_0\to -\infty$), and $\Gamma_0\to \infty$ with
$\Gamma(t)=-vt$ for every $t$, the problem can be solved analytically
(in terms of parabolic cylinder functions, see for instance Ref.\ \onlinecite{QA_jpn} for
the RT case) for both RT and IT. 
The probability $P_{ex}(0)$ of ending into the excited state can be 
expressed in terms of the variable $\gamma=\Delta^2/4v$. 
The explicit expressions, in terms of Gamma functions, are:
\begin{eqnarray} \label{lz_2:eqn}
P_{ex}(0) &=& \frac{|R+1|^2}{2(1+|R|^2)} \\
R &=& e^{i\phi_0} \frac{1}{\sqrt{\gamma}} \frac{{\mathbf \Gamma}(1+z)}
                                             {{\mathbf \Gamma}(1/2+z)} \nonumber \;,
\end{eqnarray}
where $\phi_0=3\pi/4$ and $z=i\gamma$ for RT, while $\phi_0=\pi$ and $z=\gamma$ for IT. 
%Here ${\mathbf \Gamma}$ is the Gamma function. 
A plot of $P_{ex}$ for both RT and IT is shown in Fig.\ \ref{zener:fig} as a function of
$\gamma=\Delta^2/4v$.
%
%------------------------------------------------------------------------
%------------------------------------------------------------------------
%------------------------------------------------------------------------
\begin{figure}[!tbp]
  \includegraphics*[width=8cm,angle=0]{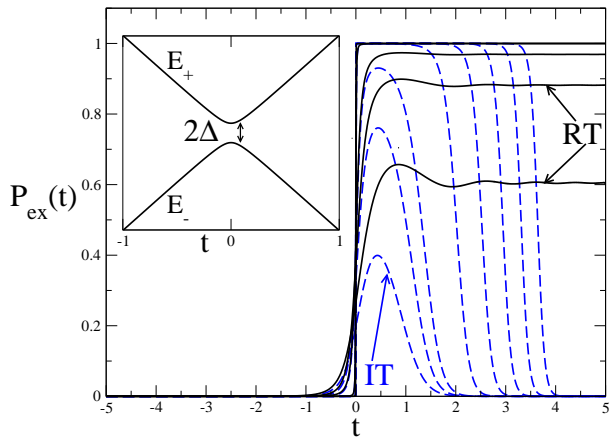}
  \caption{\label{comp_IT_RT:fig} 
   Comparison between the RT (solid lines) and the IT (dashed lines) evolution 
   of a Landau-Zener problem, Eq.~\protect\ref{lz_rot:eqn}, 
   for several values of the tunneling gap $2\Delta$ (the values of $\Delta$
   shown are $\Delta=0.4,0.2,0.1,10^{-2},10^{-3},10^{-4},10^{-5},10^{-6}$, 
   while $v=1$). 
   The inset shows the two instantaneous eigenvalues of the problem, 
   $E_{\pm}(t)$, as a function of $t$.
   }
\end{figure}
%------------------------------------------------------------------------
%
Note that: i) the IT-result for $P_{ex}$ (dashed line) is always below 
the RT-result, ii) the difference between the two curves is only quantitative: 
one can verify analytically that the leading behavior for large $\gamma$ is the
same in both cases, i.e., $P_{ex}\approx 1/(256\gamma^2)$. 
Similar results are obtained by direct numerical integration of the
Scr\"odinger equation for finite $\Gamma_0$ and $\tau$, and with 
other forms of $\Gamma(t)$. 

With the same toy model, we can illustrate another point raised in the previous
Section: what happens to the IT evolution {\em after} a Landau-Zener avoided crossing
gap is encountered. The Hamiltonian we consider is essentially that in Eq.~\ref{lz_1:eqn},
simply rotated in spin space, 
\begin{equation} \label{lz_rot:eqn}
H(t) = -vt \, \sigma^z - \Delta \sigma^x \;.
\end{equation}
In absence of the tunneling amplitude $\Delta$, the two energy levels would cross at $t=0$,
while for $\Delta>0$ the two instantaneous eigenvalues are simply 
$E_{\pm}(t)=\pm \sqrt{(vt)^2+\Delta^2}$ (see inset in Fig.~\ref{comp_IT_RT:fig}). 
Starting with the system in the ground state at 
$t=-\infty$, we can monitor the probability of getting onto the excited states at any time
$t$, which we plot in Fig.~\ref{comp_IT_RT:fig} for both the RT and the IT evolution and
for several values of $\Delta$ (taking $v=1$). 
The RT data provide an illustration of the well-known Landau-Zener result: 
after a (relatively short) tunneling time, and possibly a few oscillations, the probability of
of getting onto the excited state saturates to a value given by 
$P_{ex}(t=\infty)=e^{-\pi\Delta^2/\hbar v}$. 
As for the IT data, the initial (tunneling) part and the subsequent plateau of the 
curves are similar to the RT case: the plateau value attained, call it $P_{ex}^*$, 
is indeed very close to the RT saturation value (in fact, asymptotically the same 
for $\Delta\to 0$); 
after that, the IT evolutions starts to filter out the ground state component 
-- initially present in the state with a small amplitude $1-P_{ex}^*$ -- through 
the usual mechanism of suppression of excited states, leading to a $P_{ex}(t)$ which
is nicely fit by the curve
\begin{equation} 
P_{ex}(t) = \frac{ P_{ex}^* e^{-2\int_0^t dt' [E_+(t')-E_-(t')] }                }
                 { (1-P_{ex}^*) + P_{ex}^* e^{-2\int_0^t dt' [E_+(t')-E_-(t')] } } \;,
\end{equation}
which asymptotically goes to zero as $t\to \infty$.
This rather trivial effect of filtering, if on one hand explains the discrepancy between
the IT and the RT evolution observed in the asymmetric double well case of the previous section,
is, on the other hand, of no harm at all: on the contrary, it provides a quantitative improvement
of IT over RT. 

In summary, the essential equivalence of IT and RT Schr\"odinger annealing (with, moreover,
a quantitative improvement of IT over RT) justifies practical implementations of 
quantum annealing based on imaginary-time Quantum Monte Carlo schemes. 

%------------------------------------------------------------------------
\section{One-dimensional curved washboard: a potential with many minima}
\label{washboard:sec}
%------------------------------------------------------------------------

After discussing at length the annealing problem in a potential with one minimum and
with two minima, we wish to move on to a multi-minima problem, however simple. 
There are simple but interesting one-dimensional potentials which
allow us to do that.
The first example was proposed and solved by Shinomoto and Kabashima 
in Ref.\ \onlinecite{Shi_Kab}, and consists in a parabolically shaped washboard 
potential. This example will display a logarithmically slow classical annealing, 
showing CA may run into trouble even in simple models with no complexity
whatsoever, whereas quantum mechanics can do much better in this case.
%
%------------------------------------------------------------------------
\begin{figure}[!tbp]
  \includegraphics*[width=8cm,angle=0]{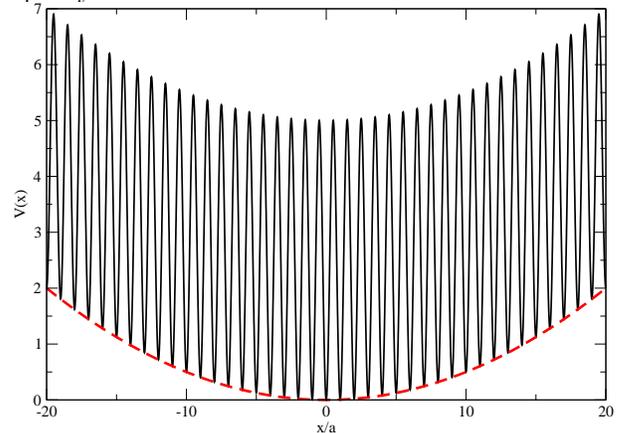}
  \caption{\label{wigged:fig} 
    Parabolic washboard potential resulting in a logarithmically slow 
    classical annealing. The minima are regularly located at positions
    $x_i=ia$, and the dashed line shows the parabolic envelope potential.
    }
\end{figure}
%------------------------------------------------------------------------
%
Consider a wiggly one-dimensional potential with barriers of individual height
$\approx B$ separating different local minima, regularly located a distance $a$ 
apart one from each other, i.e., at positions $x_i=ai$. The ith-local minimum 
is at energy $\epsilon_i=ka^2i^2/2$, so that the resulting envelope is parabolic.
In order to study the dynamics of a particle in this potential,
a good starting point is to write the master equation for the probability
$P_i(t)$ that the particle is in the ith-valley at time $t$:
\begin{widetext}
\begin{equation} \label{master:eqn}
\frac{1}{\gamma} \frac{d}{dt} P_i(t) =
P_{i+1}(t) e^{-\beta B_{i+1,i}} +
P_{i-1}(t) e^{-\beta B_{i-1,i}} -
P_i(t)\left[ e^{-\beta B_{i,i+1}} + e^{-\beta B_{i,i-1}} \right] \;,
\end{equation}
\end{widetext}
where $\gamma$ is an attempt frequency, $B_{i,j}$ is the
effective barrier from $i$ to $j$, and $\beta=1/k_BT$.
This is a well justified starting point, in view of the results of the
previous sections (Secs.\ \ref{tls:sec} and \ref{harmonic:sec}), showing 
that classical annealing is extremely fast
in reducing the width of a probability distribution within each valley
to an essentially delta-function like sequence of peaks of strength $P_i$.
The actual form of the effective barriers depends on the way we model
the details of the potential, with the only constraint that detailed
balance is satisfied, i.e.,
\[ B_{i,i+1}-B_{i+1,i} = \epsilon_{i+1}-\epsilon_i = \Delta_i \;, \]
in such a way that the stationary solution, for constant $T$, is simply 
the Boltzmann probability distribution, 
$P_i(t\to\infty) \propto exp(-\epsilon_i/k_BT)$.
Ref.\ \onlinecite{Shi_Kab} takes $B_{i+1,i}=B_{i,i-1}=B$, while
$B_{i,i+1}=B+\Delta_i$ and $B_{i-1,i}=B+\Delta_{i-1}$, with
$\Delta_i = \epsilon_{i+1}-\epsilon_i$. 
%
%Notice also the following important point: 
The potential energy of the valleys enters only through the $B_{i,j}$, 
which control the probability of making transitions between valleys. 
 
In order to study Eq.~(\ref{master:eqn}), Shinomoto and Kabashima introduced 
a continuum limit, by defining a macroscopic coordinate $x$, such that the 
mimima are at $x_i=ia$, and writing the equation governing the probability 
$P(x,t)$ in the limit $a\to 0$.
The derivation involves writing $P_{i\pm 1}(t)$ in terms of derivatives of 
$P(x,t)$, keeping consistently terms up to order $a^2$ and expanding 
exponentials with the assumption that $k_BT/(ka^2)>>1$.
The continuum limit equation governing the evolution of $P(x,t)$ turns out 
to be a Fokker-Planck (FP) equation, Eq.~(\ref{FP:eqn}), with
an effective diffusion constant of the form
\begin{equation} \label{Diff_const:eqn}
D_{\rm eff}(T) = \gamma a^2 e^{-B/k_BT} \;,
\end{equation}
$\eta(T)=k_BT/D_{\rm eff}(T)$, and an effective drift potential $V(x)=kx^2/2$ 
given by the macroscopic parabolic envelope potential. 
In order to study the annealing properties of the system, one can then follow 
exactly the same steps leading to Eq.~(\ref{der_residual_closed:eqn}), which 
applies here too, except that now $D_t$ is substituted by 
$D_{\rm eff}(T(t))$, which has an exponential activated behavior,
$D_{\rm eff}(T)\propto e^{-B/k_BT}$. 
This exponentially activated $D_{\rm eff}(T)$ changes the annealing behavior 
in a drastic way. 
Recall that the CA exponent $\Omega_{CA}$ of Sec.~(\ref{harmonic:sec})
decreases towards zero as the exponent $\alpha_D$ in the relationship
$D(T)\propto T^{\alpha_D}$ increases. 
Since, close to $T=0$, $e^{-B/k_BT}<< T^{\alpha_D}$ for any arbitrarily
large $\alpha_D$, we could suspect that the behavior of
$\epsilon_{res}(\tau)$ will no longer be a power law.
In fact, the surprising result of this exercise \cite{Shi_Kab} is that the 
optimal annealing schedule $T(t)$ is {\em logarithmic} and that 
$\epsilon_{res}(t)$ converges to $0$ at best as \cite{log_continuum:nota}
\begin{equation} \label{eps_res_log:eqn}
\epsilon_{res}(t) \sim \log(t)^{-1} \;.
\end{equation}
The reason behind this result is that the time derivative of 
$\epsilon_{res}(t)$ becomes exponentially small as one anneals $T$ 
towards $0$, due to the presence of $e^{-B/k_BT}$ in the diffusion constant, 
and an exponentially small derivative brings -- not surprisingly -- 
a logarithmically slow decrease of the function.
To put it more physically, consider solving 
Eq.~( \ref{der_residual_closed:eqn}) for a time independent $T$; the solution 
is trivially 
\[ 
\epsilon_{res}(t) = C e^{-t/t_{\rm relax}} + \frac{k_BT}{2} \hspace{10mm} 
t_{\rm relax} = \frac{k_BT}{2\gamma ka^2} e^{B/k_BT} \;.
\]
We observe that the solution converges to the equilibrium (equipartition) 
value $k_BT/2$ exponentially with a characteristic time $t_{\rm relax}$ 
which itself increases exponentially fast with decreasing $T$. 
As a result, the system will never be able to follow the decreasing $T$ 
till the end of the annealing, by maintaining roughly the 
equilibrium value $\epsilon_{pot}=k_BT/2$. 
Indeed, if we assume for instance $T(t)=T_0(1-t/\tau)$, the relaxation of 
the systems will cease to be effective 
-- i.e., the system will fall {\em out of equilibrium} -- at a time $t^*$, 
and temperature $T^*=T(t^*)$, at which $t_{\rm relax}\approx \tau$, 
i.e., when $k_BT^* \approx B/\log{\gamma\tau}$. 
The residual energy at this point cannot be smaller then the equipartition 
value $k_BT^*/2$, hence $\epsilon_{res} \approx B/\log{\gamma\tau}$ as well.
This freezing and falling out of equilibrium for classical systems with
barriers seems to provide an ubiquitous source of logarithms in classical 
annealing \cite{Huse_Fisher}.

How would one tackle this annealing problem quantum mechanically?  
As the quantum analog of the master equation Eq.~(\ref{master:eqn}), we
propose studying the Schr\"odinger evolution governed by a tight-binding 
Hamiltonian with on-site energies $\epsilon_i$ and hopping matrix-elements 
between adjacent sites $h_{i,i+1}$ 
\begin{equation} \label{H_tb:eqn}
H = \sum_i \epsilon_i c^{\dagger}_i c_i - 
\sum_i h_{i,i+1}\left( c^{\dagger}_i c_{i+1} + c^{\dagger}_{i+1} c_{i}  \right) \;.
\end{equation}
The justification and possible limitations of this starting point, over the actual 
original continuum problem will be discussed at the end of the section.
Here it is enough to consider that the hopping matrix-elements $h_{i,i+1}$ depend 
on tunneling through the barrier separating $i$ from $i+1$. 
The precise form of $h_{i,i+1}$ is likely to be inessential, including its energy
(and hence site) dependence, for which we will assume the semi-classical (WKB) form:
\begin{equation} \label{t_tb:eqn}
h_{i,i+1} \sim h = h_0 \left( \frac{V_h}{\Gamma} \right)^{1/4} 
e^{-\sqrt{V_h/\Gamma}} \;,
%\hspace{12mm} \mbox{with} \; \Gamma = \frac{\hbar^2}{2ma^2} \;,
\end{equation}
$\Gamma=\hbar^2/(2ma^2)$ being simply related to the quantum confinement 
energy of a particle of mass $m$ in a valley of size $\sim a$.
Here $h_0$ and $V_h$ are energy parameters related to the details of the 
potential and of the barrier, which will play little or no role. 
If the mass of the particle $m$ (and hence $\Gamma$ and $h_{i,i+1}$)
is kept constant the particle will explore the potential due to the kinetic
term in the Hamiltonian: the correspondence between the quantum and the
classical formulation is that $\Gamma$ plays the role of $T$, 
$h_{i,i+1}$ plays the role of the classical transition probabilities, the
ground state wavefunction $|\Psi_i^{(\Gamma,GS)}|^2$ at a given value of 
$\Gamma$ (or, equivalently, of the hopping term $h$) plays the role of the 
classical equilibrium Boltzmann distribution $P_i^{(T,eq)}$. 
The question, once again, is how to anneal $\Gamma$, by reducing
it as a function of time, $\Gamma(t)$, in such a way as to squeeze 
the wavefunction of the system so that the average potential energy
\begin{equation}
\epsilon_{pot}(t)=\frac{\sum_i \epsilon_i |\Psi_i(t)|^2}{\sum_i |\Psi_i(t)|^2} 
\end{equation}
is minimal.

As it turns out, the continuum limit is once again useful. One goes to the
continuum exactly as in the FP case by using $x_i=ai$, writing 
$\psi(x_i,t)=\Psi_i(t)/\sqrt{a}$ and expanding everything to order $a^2$. 
When written in first quantized form, the Hamiltonian for the
quantum particle in the macroscopic continuum coordinate $x$ is simply
\begin{equation}
H(t) = -\Gamma_{\rm eff}(t) \nabla^2 + V(x) 
\end{equation}
where the coefficient of the Laplacian $\Gamma_{\rm eff}$ is the quantum 
counterpart of the classical effective diffusion constant $D_{\rm eff}$
\begin{equation} \label{Gamma_eff:eqn}
\Gamma_{\rm eff}(t) = a^2 h(t) = 
a^2 h_0 \left( \frac{V_h}{\Gamma(t)} \right)^{1/4} e^{-\sqrt{V_h/\Gamma(t)}} \;,
\end{equation}
and $V(x)=kx^2/2$, as in the classical case. 
The continuum limit quantum problem has therefore exactly the form we have 
considered in Sec.~(\ref{harmonic:sec}), except that $\Gamma(t)$ in 
Eq.~(\ref{Gauss_ODE_B:eqn}) is now substituted by an effective Laplacian 
coefficient $\Gamma_{\rm eff}$ which has a highly non-linear, in fact 
exponential, dependence on the annealing parameter $\Gamma(t)$. 
We know, however, from Sec.~(\ref{harmonic:sec}), that a non-linear behavior 
of the type $(1-t/\tau)^{\alpha_{\Gamma}}$ for the Laplacian coefficient 
leads to a power-law decrease of $\epsilon_{res}(\tau)$ with an exponent 
$\Omega_{QA}$ which is, remarkably, an 
{\em increasing function of $\alpha_{\Gamma}$}, 
approaching $1$ as $\alpha_{\Gamma}\to \infty$.
Therefore, contrary to the classical case, where an exponentially activated
behavior of the diffusion constant $D_{\rm eff}$ is 
{\em strongly detrimental} to the annealing (turning a power law into a 
logarithm), here the exponential WKB-like behavior of $\Gamma_{\rm eff}$ 
will do no harm at all. 
Indeed, we numerically integrated the relevant equation for $B_t$, 
Eq.~(\ref{Gauss_ODE_B:eqn}) with $\Gamma_{\rm eff}$ in place of $\Gamma$, 
using the exponential WKB expression Eq.~(\ref{Gamma_eff:eqn}) for 
$\Gamma_{\rm eff}$ while annealing $\Gamma(t)$ with a linear schedule, 
$\Gamma(t)=\Gamma_0(1-t/\tau)$. 
The integration was performed, as usual, with a fourth-order Runge-Kutta 
method, and was carried on up to time $t=\tau$, when the kinetic 
term in the Hamiltonian ceases to exist. 
The numerical results (not shown) have a clear power-law behavior for the 
quantum annealed (QA) final residual energy 
$\epsilon_{res}(t=\tau) \sim \tau^{-\Omega_{QA}}$, 
with a power law exponent $\Omega_{QA}$ which is compatible with $1$.
Once again, the exponent appears to be
insensitive to the choice of the type of quantum evolution (RT versus IT), 
although the numerical values of residual energies always respect the
inequality $\epsilon_{res}^{RT}(\tau) \ge \epsilon_{res}^{IT}(\tau)$. 

Before ending this section, we would like to discuss briefly the reason for
treating by tight-binding, Eq.\ \ref{H_tb:eqn}, what was originally
a continuum problem with a well defined potential landscape.
As we learned from the double-well case, there is never a clear-cut 
discrete model (a discrete two-level system, in that case) describing in a complete
way the continuum Schr\"odinger problem, in all stages of the annealing.
Obviously, when the mass of the particle is very small, the tight-binding 
approximation contained in Eq.\ \ref{H_tb:eqn} is not particularly good, since more
than one state per valley is generally important to describe the wavefunction
accurately. 
As the mass of the particle increases, however, the tight-binding approach gets more 
and more appropriate, until a further limitation appears: when the mass if very 
large, it is not legitimate to neglect excited states within, say, the central
valley compared to the lowest states localized in metastable valleys.
We can imagine that the ultimate behavior of $\epsilon_{res}(\tau)$, in the
quantum case, will be actually dominated by the rather trivial problem of
squeezing the wavefunction in the lowest central minimum, with its characteristic
power-law exponent ($1/3$, for instance, for a linear schedule). 
There is, however, an intermediate region, between the very short $\tau$ scale, where
the full details of the potential are important, and the very long $\tau$ scale,
where the trivial squeezing mentioned above sets in, and where the tight-binding 
approximation reasonably predicts a power-law exponent (of order $1$) for
$\epsilon_{res}(\tau)$. 

We believe that one of the important points that makes QA so different from CA in the
present case is that the spectrum of the instantaneous eigenvalues of the quantum problem
does not show any dangerous Landau-Zener avoided-crossing, 
and, correspondingly, the ground state
wavefunction is always more peaked in the central valley than elsewhere.
As in the two-level case, a disorder in the width of the different valleys would change 
this result.

%------------------------------------------------------------------------
\section{Role of Disorder}
\label{disorder:sec}
%------------------------------------------------------------------------

\begin{figure}
  \includegraphics*[width=8cm,angle=0]{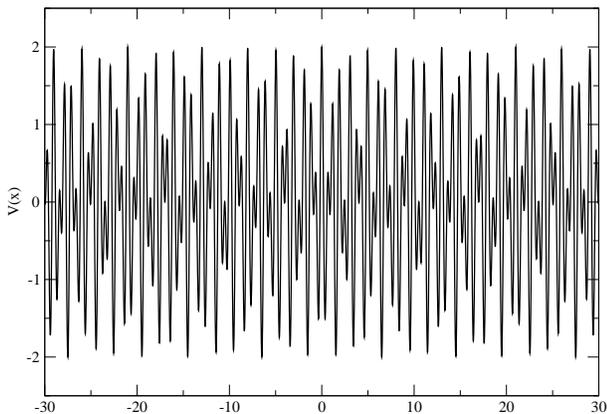}
  \label{double_cos:fig}
  \caption{Double cosine potential $V(x)=cos(2\pi x)+cos((1+\sqrt{5})\pi x)$,
           showing an irregular landscape with many minima.}
\end{figure}

Despite their disarming simplicity, the three case studies above turn out to
be extremely informative in qualifying the profound difference of QA from CA,
and their surprising consequences. We expect that these results will be very
important in understanding more realistic QA problems.
Of course, the cases studied, although instructive, do not possess the 
real ingredient which makes annealing difficult, both in CA and QA, i.e., some
form of disorder in the distribution of the minima. 
We believe, for instance, that even an irregular landscape with many minima, 
as the double-cosine potential $V(x)=V_1\cos{(2\pi x)}+V_2\cos{(2\pi r x)}$
(with $r$ an irrational number) shown in Fig.\ \ref{double_cos:fig}, 
would already change drastically the behavior 
of QA from a power-law to a logarithm.
On quite general grounds, Anderson's localization \cite{and_loc:rev}
would predict that wavefunctions 
are localized for a genuinely disordered potential for large enough mass (i.e., small
enough kinetic energy bandwidth) in any $D>2$ (this localization occurs for all value
of the mass in $D=1,2$). 
Therefore, quantum annealing should always, via a cascade of Landau-Zener 
events, end up into some localized state which has, a priori, nothing to do 
with our search of the actual potential minimum.
%
%------------------------------------------------------------------------
%------------------------------------------------------------------------

A very simply illustration of the crucial role of disorder is given by the $D=1$
disordered Ising ferromagnet:
\begin{equation}
H=-\sum_i J_i \sigma^z_i \sigma^z_j - \Gamma \sum_i \sigma^x_i
\end{equation}
where $J_i\ge 0$ are positive random variables in the interval $[0,1]$, and $\Gamma$ is
the transverse field inducing quantum fluctuations.
Obviously, the ground state is the ferromagnetic state with all spins aligned up (or down).
However, arbitrarily weak values of the $J_i$ can pin domain walls between up and down
ferromagnetic regions, with a very small energy cost $2J_i$. 
For a finite system with periodic boundary conditions domain walls appear in pairs, and
separate sections of the system with alternating $\uparrow$ and $\downarrow$ ferromagnetic ground
states. Given two domain walls pinned at weak-$J_i$ points a distance $L\gg 1$ apart, 
healing the system via single spin flip moves requires flipping $L$ spins, which can
be a formidable barrier to tunnel through. 
The system will have a very slow annealing (quantum, as well as classical) while showing,
at the same time, no complexity whatsoever: simple disorder is enough. 

%------------------------------------------------------------------------
\section{Summary and conclusions}
\label{summary:sec}
%------------------------------------------------------------------------
Summarizing, we have compared Schr\"odinger versus Fokker-Planck annealing in
various simple cases of one-dimensional potential, in particular a double well
potential and a parabolic washboard potential with many minima but no disorder.
In all cases the two annealings, quantum and classical, were seen to behave in 
a remarkably different way. 
Classical annealing is influenced only by the height of the barriers surrounding 
the relevant minima, via Arrhenius-like thermal promotion over the barriers, with 
probability distributions which are quite localized on those minima. 
Quantum annealing is influenced by the structure of the eigenvalue spectrum of the 
problem -- very small Landau-Zener tunneling gaps associated to large barriers
are highly detrimental to it. 
In some cases, for instance in the case of the parabolic washboard potential, quantum
annealing can be much more effective then classical annealing, but, generally speaking, 
both strategies suffer whenever the potential landscape is disordered.
As an outcome of the discussion, it is quite clear that quantum annealing, although
potentially useful and sometimes more convenient than classical annealing, is not
capable, in general, of finding solutions of NP-complete problems in polynomial time:
indeed, quite interestingly, even trivial optimization problems (trivial, obviously, only
from the complexity point of view), like the one-dimensional ferromagnetic random Ising model,
can show a very slow annealing behavior.

%--------------------- ACKNOWLEDGEMENTS ---------------------------------
\begin{acknowledgments} 
This project was sponsored by MIUR through FIRB RBAU017S8R004, FIRB RBAU01LX5H, 
COFIN2003 and COFIN2004, and by INFM (``Iniziativa trasversale calcolo parallelo''). 
We acknowledge illuminating discussions with Demian Battaglia.
\end{acknowledgments}

%------------------------------------------------------------------------
% APPENDIX
%------------------------------------------------------------------------
\appendix

\section{Annealing in a parabolic potential}

\subsection{Classical Annealing (Fokker-Planck) case.}
%{\bf Classical Annealing (Fokker-Planck) case.}
%
If is straightforward to find the solution of the Fokker-Planck 
Eq.~(\ref{FP:eqn}) when the the potential is harmonic, $V(x)=kx^2/2$, and 
the initial condition is the Boltzmann distribution 
$P(x,t=0)\propto \exp{(-kx^2/(2k_BT_0))}$. Indeed, it is simple to
verify that the Gaussian {\it Ansatz} 
\begin{equation} \label{Gauss_FP:eqn}
P(x,t) = C_t e^{-B_tx^2}  \;,
\end{equation}
fulfills the initial condition (if $B_{t=0}=B_0=k/(2k_BT_0)$) and solves 
the FP equation, as long as the two functions $B_t$ and $C_t$ satisfy the 
following ordinary differential equations:
\begin{equation} \label{Gauss_ODE_FP:eqn}
\left\{ \begin{array}{l}
\dot{B}_t = 2D_t \left( \frac{kB_t}{k_BT(t)} -2B_t^2 \right) \\
\dot{C_t}/C_t = D_t \left( \frac{k}{k_BT(t)} -2B_t \right) \\
\end{array}  \right.  \;.
\end{equation}
The normalization constant $C_t$ turns out to be irrelevant in calculating 
the average potential energy which we need
\begin{equation} \label{E_pot_FP:eqn}
\epsilon_{pot}(t) = \frac{\int dx V(x) P(x,t)}{\int dx P(x,t)} =
\frac{k}{4B_t} \;,
\end{equation}
and can be completely forgotten, since the equation for $B_t$ does not 
involve it. The equation for $B_t$ appears to be {\em non-linear}, but
can immediately transformed into a linear equation by dividing up both sides 
by $B_t^2$ and recognizing that the correct variable to use is precisely 
$1/B_t$, or better yet, $\epsilon_{pot}(t)$. 
In terms of $\epsilon_{pot}(t)$ we can therefore write a linear equation 
of the form: 
\begin{equation} \label{der_E_pot:eqn}
\frac{d}{dt} \epsilon_{pot}(t) = 
k D_t \left( 1 - \frac{2}{k_B T(t)} \epsilon_{pot}(t) \right) \;, 
\end{equation}
the initial condition being simply given by the equipartition value
$\epsilon_{pot}(t=0)=(k/4B_0)=k_BT_0/2$. 
An alternative way \cite{Shi_Kab} of deriving Eq.~(\ref{der_E_pot:eqn}) 
consists in taking the derivative with respect to time of both sides of 
Eq.~(\ref{E_pot_FP:eqn}), 
%
%\begin{equation} \label{der_residual:eqn}
%\frac{d}{dt} \epsilon_{pot}(t) = 
%\int dx \, \left( \frac{k}{2} x^2 \right) \, 
%\frac{\partial}{\partial t} P(x,t) \;,
%\end{equation}
%
using then the FP equation for $\partial P/\partial t$ on the left hand-side 
of the ensuing equation, and finally integrating by parts the terms containing 
spatial derivatives of $P$ (this procedure results in a closed differential 
equation for $\epsilon_{pot}(t)$ only if the potential $V(x)$ is harmonic).
As every one-dimensional linear first-order differential equation, 
Eq.~(\ref{der_E_pot:eqn}) can be integrated by quadrature, the solution 
being:
\begin{eqnarray}
\epsilon_{pot}(t) &=& \epsilon_{pot}(t=0) \, e^{-(k/2k_B)\int_0^t dy D_y/T(y)} 
\nonumber \\
&& + k \int_0^t dt' D_{t'} e^{-(k/2k_B)\int_{t'}^t dy D_y/T(y)} \;.
\end{eqnarray}
If we now anneal the temperature down to $0$ in a time $\tau$ in the usual 
way, $T(t)=T_0(1-t/\tau)^{\alpha_T}$, assuming that
the diffusion constant behaves as $D(T)=D_0(T/T_0)^{\alpha_D}$, 
we readily get an analytic expression for the residual energy at the end
of the annealing, $\epsilon_{res}(\tau)=\epsilon_{pot}(t=\tau)$, 
which can be shown to behave, for large $\tau$, as a power law:
\begin{equation} \label{CA_expon:eqn}
\epsilon_{res}(\tau) \approx \tau^{-\Omega_{CA}} \hspace{12mm}
\Omega_{CA} = \frac{\alpha_T}{\alpha_T(\alpha_D-1)+1} \;.
\end{equation}
Quite evidently, annealing proceeds here extremely fast, with a power-law 
exponent $\Omega_{CA}$ that can increase without bounds (for instance 
if $\alpha_D=1$) upon increasing the exponent $\alpha_T$ of the annealing
schedule. 
Notice, however, that large values of $\alpha_D$ are, on the contrary, 
detrimental.

\subsection{Quantum Annealing (Schr\"odinger) case.}
Consider now the problem of a particle moving in a parabolic potential 
$V(x)=kx^2/2$, with a time-dependent mass, such that the Hamiltonian is
given by:
\begin{equation}
H(t) = \frac{k}{2} x^2 - \Gamma(t) \nabla^2 \;,
\end{equation}
where $\Gamma(t)=\hbar^2/2m(t)$ denotes the coefficient of the Laplacian
operator.
The Schr\"{o}dinger evolution of the wavefunction $\psi(x,t)$ is then, 
\begin{equation} \label{Schr_evol:eqn}
\xi \partial_t \psi(x,t) = H(t) \psi(x,t) \;, 
\end{equation}
where $\xi=i\hbar$ for a real time (RT) evolution, and 
$\xi=-\hbar$ for an imaginary time (IT) evolution.
Now, whereas general solutions of the time-dependent Schr\"{o}dinger equation 
for arbitrary initial condition $\psi(x,t=0)$ are not easy, it turns out
that if $V(x)$ is quadratic, then any initial Gaussian wavefunction 
propagates into a Gaussian, which is enough for our goal. 
In detail, write the following {\it Ansatz} for $\psi(x,t)$: 
\begin{equation} \label{Gauss:eqn}
\psi(x,t) = C_t e^{-B_tx^2/2}  \hspace{12mm} {\rm Real}(B_t)=\Re(B_t) > 0 \;.
\end{equation}
Substituting the {\it Ansatz} for $\psi(x,t)$ into the Schr\"{o}dinger 
evolution (in RT or in IT), one immediately verifies that $\psi(x,t)$ satisfies 
Eq.~(\ref{Schr_evol:eqn}) for arbitrary $\Gamma(t)$ as long as $B_t$ and $C_t$ 
satisfy the following ordinary differential equations:
\begin{equation} \label{Gauss_ODE:eqn}
\left\{ \begin{array}{l}
-\xi \dot{B_t} = k - 2 \Gamma(t) B_t^2 \\
-\xi \dot{C_t}/C_t = -\Gamma(t) B_t   \\
\end{array}  \right.  \;.
\end{equation}
Once again, the normalization constant $C_t$ turns out to be irrelevant in 
calculating the average potential energy 
\begin{equation}
\epsilon_{pot}(t) = \frac{\int dx V(x) |\psi(x,t)|^2}{\int dx |\psi(x,t)|^2} =
\frac{k}{4\Re(B_t)} \;,
\end{equation}
and can be completely forgotten, since the equation for $B_t$ does not involve
it. 
The initial condition $B_{t=0}$ is, by assumption, such that at $t=0$ the 
system is in the {\em ground state} corresponding to $\Gamma_0=\Gamma(t=0)$.
Such a ground state value $B_0$ is easily calculated by equating to zero 
the right-hand side of Eq.~(\ref{Gauss_ODE:eqn}) with $\Gamma(t)\to \Gamma_0$, 
i.e., $B_0=\sqrt{k/(2\Gamma_0)}$.

A few general considerations can be based on a purely qualitative 
analysis of Eq.~(\ref{Gauss_ODE:eqn}).
Consider first the IT case, where the equation for $B_t$ reads 
$\hbar\dot{B_t} = k - 2 \Gamma(t) B_t^2$. 
One can easily get convinced that $B_t$ is forced to be a real, positive and 
monotonically increasing function of $t$, i.e., $B_t>0$ and $\dot{B_t}>0$ for 
$t>0$. Therefore, we can easily write an inequality of the form:
\[ \hbar\dot{B_t} = k - 2 \Gamma(t) B_t^2 \le k \;, \]
from which we immediately conclude, by integrating over $t$ the two
sides of the inequality, that 
\[ B_{\tau} - B_0 \le \frac{k\tau}{\hbar} \;, \]
i.e., the residual energy $\epsilon_{res}(\tau)=k/(4B_{\tau})$ cannot
decrease faster than $1/\tau$ for $\tau\to \infty$.

We will now assume, without great loss of generality, that the Laplacian 
coefficient $\Gamma(t)$ is given by $\Gamma(t)=\Gamma_0 f(t/\tau)$, where 
$\tau$ is an annealing time-scale (for instance the annealing time, when 
a linear schedule is used) and $f(t')$ is a positive decreasing function 
for $t'\ge 0$ such that $f(t'\le 0)=1$. 
It is useful to switch to dimensionless variables by measuring times in
unit of $\tau$, $t'=t/\tau$, and $B_t$ in units of its initial ground state
value $B_{t=0}=B_0$.
The appropriate dimensionless quantity is therefore $b(t';\tau)=B_t/B_0$,
with $t'=t/\tau$, where the parametric dependence on the annealing time-scale 
$\tau$ has been explicitly indicated. %but will be at times omitted. 
The equation for $b(t';\tau)$ is given by:
\begin{eqnarray} \label{non-linear:eqn}
\dot{b}(t';\tau) &=& \alpha [1 - f(t') b^2(t';\tau)] 
\qquad \alpha = \tau \sqrt{2 k \Gamma_0 }/(-\xi) \nonumber \\ 
b(0;\tau) &=& 1  \;,
\end{eqnarray}  
where the dot, from now on, will denote a derivative with respect to $t'$.
Notice that the parametric dependence on $\tau$ is all buried into the
coefficient $\alpha$, which reabsorbs also the $-\xi$ appearing in the
dynamics (RT versus IT).
This kind of non-linear differential equation is of the well known 
{\em Riccati} form. 
It can be transformed into a \emph{linear second-order} differential equation 
by operating the following substitution
\begin{equation} \label{b_versus_y:eqn}
b(t';\tau) = \frac{\dot{y}(t';\tau)}{\alpha f(t') y(t';\tau)} \;,
\end{equation}
where, evidently, $y$ is defined up to an overall normalization constant.
Indeed, simple algebra shows that we can re-express Eq.~(\ref{non-linear:eqn})
as a second-order linear equation for $y$, as follows:
\begin{eqnarray} \label{linear_1:eqn}
&& f(t') \ddot{y}(t';\tau) - \dot{f}(t') \dot{y}(t';\tau) 
= \alpha^2 f^2(t') y(t';\tau) \hspace{2mm} \\
&& \dot{y}(0;\tau) = \alpha y(0;\tau) \;. \nonumber 
\end{eqnarray}
As long as $f(t')\ne 0$, it is simple to verify that the second-order
equation for $y(t';\tau)$ can be also equivalently written as:
\begin{equation} \label{linear_2:eqn}
\frac{d}{dt'} \left( \frac{\dot{y}(t';\tau)}{f(t')} \right) 
= \alpha^2 y(t';\tau) \;.
\end{equation}
Finally, denoting by $Y(t';\tau)$ the indefinite integral of $y(t';\tau)$,
such that $\dot{Y}=y$, and integrating over $t'$ both sides of 
Eq.~(\ref{linear_2:eqn}), we can also write:
\begin{equation} \label{gen_Airy:eqn}
\ddot{Y}(t';\tau) = \alpha^2 f(t') Y(t';\tau) \;.
\end{equation}
Eq.~(\ref{gen_Airy:eqn}) is easily solved when the annealing schedule for
$\Gamma(t)$ is linear, $\Gamma(t)=\Gamma_0 (1-t/\tau)$, i.e., when 
$f(t')=(1-t')^{\alpha_{\Gamma}}$ with $\alpha_{\Gamma}=1$, a case in 
which Eq.~(\ref{gen_Airy:eqn}) is of the {\em Airy} type.
In the latter case, it is useful to perform a final change of independent
variable to $z=\xi^2|\alpha|^{2\over 3}(1-t')$, so that, defining 
$F(z)=Y(t';\tau)$, we can write the equation for $F$ in the standard
Airy form:
\begin{equation} \label{Airy:eqn}
\frac{d^2}{dz^2} F(z) = z F(z) \;.
\end{equation} 
The general solution of Eq.~(\ref{Airy:eqn}) is given in terms of the two 
Airy's functions $Ai(z)$ and $Bi(z)$, 
\begin{equation}
F(z) = \beta Ai(z) + \gamma Bi(z) \;,
\end{equation}
where $\beta$ and $\gamma$ are two constant coefficients.
Going back to $Y(t';\tau)$ and $y(t';\tau)$, we then have the explicit 
expressions:
\begin{widetext}
\begin{eqnarray} \label{solution_Y:eqn}
Y(t';\tau) &=&  
  \beta Ai(\xi^2|\alpha|^{2\over 3}(1-t')) + \gamma Bi(\xi^2|\alpha|^{2\over 3}(1-t')) \nonumber \\
\dot{Y}(t';\tau) &=& y(t';\tau) = 
 -\beta  \xi^2|\alpha|^{2\over 3} Ai^{\prime}(\xi^2|\alpha|^{2\over 3}(1-t')) 
 -\gamma \xi^2|\alpha|^{2\over 3} Bi^{\prime}(\xi^2|\alpha|^{2\over 3}(1-t')) \nonumber \\
\ddot{Y}(t';\tau) &=& \dot{y}(t';\tau) = 
 \beta  |\alpha|^{4\over 3} Ai^{\prime\prime}(\xi^2|\alpha|^{2\over 3}(1-t')) + 
 \gamma |\alpha|^{4\over 3} Bi^{\prime\prime}(\xi^2|\alpha|^{2\over 3}(1-t')) 
 \nonumber \\
&=& \alpha^2 (1-t') \left\{ \beta  Ai(\xi^2|\alpha|^{2\over 3}(1-t')) +
                            \gamma Bi(\xi^2|\alpha|^{2\over 3}(1-t')) \right\} \;,
\end{eqnarray}
\end{widetext}
where the prime indicates a derivative with respect to $z$, and we have used 
the property of Airy's functions that $Ai^{\prime\prime}(z)=zAi(z)$ and 
$Bi^{\prime\prime}(z)=zBi(z)$. 
Finally, substituting back the expressions in Eq.~(\ref{solution_Y:eqn})
into the original function $B_t$, see Eq.~\ref{b_versus_y:eqn}, we get:
\begin{widetext}
\begin{equation}
B_t =  B_0 \frac{\dot{y}(t/\tau;\tau)}
                  {\alpha (1-t/\tau) y(t/\tau;\tau)}
=  \xi\,B_0 |\alpha|^{1\over 3} 
  \frac{ \beta  Ai(\xi^2|\alpha|^{2\over 3}(1-t/\tau)) +
         \gamma Bi(\xi^2|\alpha|^{2\over 3}(1-t/\tau)) }
       { \beta  Ai^{\prime}(\xi^2|\alpha|^{2\over 3}(1-t/\tau)) + 
         \gamma Bi^{\prime}(\xi^2|\alpha|^{2\over 3}(1-t/\tau)) } \;.
\end{equation}
\end{widetext}
This general solution correctly depends on one parameter only,
i.e., $\gamma/\beta$, so that we can put $\beta=1$ without loss of 
generality.
We impose the initial condition $B_{t=0}=B_0$ by requiring:
\begin{equation}
\xi\,|\alpha|^{1\over 3} 
 \frac{ Ai(\xi^2|\alpha|^{2\over 3}) + \gamma Bi(\xi^2|\alpha|^{2\over 3}) }
  { Ai^{\prime}(\xi^2|\alpha|^{2\over 3}) +
    \gamma Bi^{\prime}(\xi^2|\alpha|^{2\over 3})} = 1 \;.
\end{equation}
Solving for $\gamma$, we get:
\begin{equation}
\gamma = - \frac
   { \xi^2|\alpha|^{1\over3} Ai(\xi^2|\alpha|^{2\over3}) - Ai^{\prime}(\xi^2|\alpha|^{2\over3}) }
   { \xi^2|\alpha|^{1\over3} Bi(\xi^2|\alpha|^{2\over3}) - Bi^{\prime}(\xi^2|\alpha|^{2\over3}) }
\end{equation}
Due to the asymptotic properties of the Airy's functions 
($Ai(z) \to 0$, $Ai^{\prime}(z) \to 0$ and $Bi(z) \to \infty$, 
$Bi^{\prime}(z) \to \infty$, when $z\to \infty$ on the real axis),
we conclude that for the IT case:
\begin{equation}
\gamma \to 0 \qquad \mbox{ for } \alpha \; \mbox{ (or } \tau \mbox{)} \to \infty 
\end{equation}
so that, finally,

\begin{eqnarray}
B(\tau) &=& -B_0 |\alpha|^{1\over3} 
 \frac{Ai(0)) + \gamma Bi(0)}{Ai^{\prime}(0) + \gamma Bi^{\prime}(0)} \nonumber \\
 &\simeq& 
%%- B_0 |\alpha|^{1\over3} \frac{Ai(0)}{Ai^{\prime}(0)} =
 B_0 \left( \frac{2 k \Gamma_0 \tau}{\hbar} \right)^{1\over3} 
    \frac{\Gamma({1\over3})}{3^{1\over 3}\Gamma({2\over3})} \;,
\end{eqnarray}
where we used that
$Ai(0) = {3^{-{2/3}}}/{\Gamma(2/3)}$ and 
$Ai^{\prime}(0) = -{3^{-{1/3}}}/{\Gamma(1/3)}$. 
%
%\[
%Ai(0) = \frac{3^{-{2\over3}}}{\Gamma({2\over3})}  \hspace{20mm}
%Ai^{\prime}(0) = -\frac{3^{-{1\over3}}}{\Gamma({1\over3})}  \;.
%\]
%
%L: Ho inserito questo paragrafo.
On the other hand, for the RT case we have to take the limit
$z\to -\infty$ (on the real axis) instead, and in that region all
the Airy's functions oscillate. Nevertheless is possible to show that
the value of $\gamma$ is uniformly bounded for $\tau \to \infty$.

We conclude, therefore, that for large $\tau$ and with a linear annealing
schedule, $\alpha_{\Gamma}=1$, 
\begin{equation} \label{B_tau_Airy:eqn}
B(\tau) \propto \tau^{1\over3}
\end{equation}
and, consequently, the residual energy behaves asymptotically as 
\begin{equation} \label{eps_res_Airy:eqn}
\epsilon_{res}(\tau) = \frac{k}{4\Re(B_{\tau})} \propto \tau^{-{1\over3}} \;.
\end{equation}
The generalization of this result to an arbitrary annealing exponent 
$\alpha_{\Gamma}\ge 0$ in $\Gamma(t)=\Gamma_0 (1-t/\tau)^{\alpha_{\Gamma}}$, 
is a bit more involved.  
It is however possible to establish a generalization of
Eq.~(\ref{eps_res_Airy:eqn}), for arbitrary $\alpha_{\Gamma}$,
that we checked by means of direct numerical integration.
It reads:
\begin{equation} \label{B_tau_gen:eqn}
\epsilon_{res}(\tau) = \frac{k}{4\Re(B_{\tau})} \propto 
\tau^{-\Omega_{QA}} \hspace{12mm} 
\Omega_{QA} = \frac{\alpha_{\Gamma}}{\alpha_{\Gamma}+2} \;,
\end{equation}
an expression that holds true for both RT and IT annealing.

%This prediction is in excellent agreement with the power law shown by the
%numerical integration results (Runge-Kutta type), where the exponent appears
%to be $\approx -0.343$, instead of $-1/3$. 

%-------------------------------------------------------------------------
\section{Classical annealing with quantum tools: Imaginary time Schr\"odinger evolution 
of the Fokker-Planck equation.}
\label{FP_vs_IT:sec}
%-------------------------------------------------------------------------

A side issue, but nonetheless an interesting one we wish to discuss here 
relates to classical annealing, and concerns the well-known mapping of a Fokker-Planck
equation onto an imaginary-time Schr\"odinger problem \cite{Parisi}, and its implications
on the relationship between CA and QA.
The bottom-line will be that the mapping does {\em not} imply that a FP-based CA is actually 
equivalent to QA, and moreover that such a mapping is not particularly useful in our
annealing context.

Consider, once again, the FP problem with a time-dependent temperature $T(t)$
\begin{equation} \label{FP_2:eqn}
\frac{\partial}{\partial t} P(x,t) 
\,=\, \frac{1}{\eta_t} \, {\rm div} \left( P \nabla V \right) \,+\, 
D_t \nabla^2 P \;.
\end{equation}
where both the friction coefficient and the diffusion constant are now
time-dependent quantities, which we indicate by $\eta_t=\eta(T(t))$ and $D_t=D(T(t))$.
In order to map the problem in Eq.\ \ref{FP_2:eqn} onto an imaginary-time
Schr\"odinger problem, the standard procedure \cite{Parisi} is to 
pose $P(x,t)=\Phi_0(x,t)\psi(x,t)$ and to determine $\Phi_0(x,t)$ in such a way 
as to eliminate the non-Schr\"odinger-looking drift term, turning it onto a standard
potential term. The algebra is trivial. One can show that the 
drift term is eliminated if, and only if, the $\Phi_0$ satisfies the
equation:
\[ \nabla \Phi_0(x,t) = - \frac{\nabla V}{2\eta_tD_t} \Phi_0(x,t) \;, \]
whose solution is readily found to be:
\begin{equation} 
\Phi_0(x,t) = C(t) e^{-V/(2\eta_tD_t)} \;,
\end{equation}
with $C(t)$ a function of time only, which can even be taken to be
constant without loss of generality.
By plugging $P=\Phi_0 \psi$ in the FP equation \ref{FP_2:eqn}, with $\Phi_0$ as above,
one can show that the resulting equation for $\psi(x,t)$ is indeed of
the Schr\"odinger form
\begin{equation} \label{FP_to_IT:eqn}
-\frac{\partial}{\partial t} \psi(x,t) = -D_t \nabla^2 \psi(x,t) +
V_{FP}(x,t) \psi(x,t) \;,
\end{equation}
with an effective potential $V_{FP}$ given by 
\begin{equation} 
V_{FP}(x,t) = \frac{1}{2\eta_t} \left[ \frac{ (\nabla V)^2 }{2\eta_tD_t}  -
                                       \nabla^2 V \right]
+ \frac{\partial_t \Phi_0(x,t)}{\Phi_0(x,t)} \;.
\end{equation}
The first term in $V_{FP}$ is the standard effective
potential of the Riccati form obtained in the time-independent case \cite{Parisi}.
%, $D_t \nabla^2\Phi_0/\Phi_0$,
% except for the fact that both $\eta$ and $D$ depend now on $t$. 
The second piece in $V_{FP}$ is absent in the time-independent case, and can 
be easily seen to be \cite{sign_V:nota}:
\[ \frac{\partial_t \Phi_0(x,t)}{\Phi_0(x,t)} = - V(x) \frac{d}{dt}
   \left( \frac{1}{2\eta_t D_t} \right) \;. \]
The main point we want to stress is that, by annealing $T(t)$ and hence $D_t$,
we not only reduce the coefficient of the Laplacian in Eq.\ \ref{FP_to_IT:eqn},
but we also strongly modify the {\em potential} $V_{FP}$, at variance with a genuine QA
where only the kinetic term is annealed down. The modifications of the potential are
so strong that, at low temperature, the instantaneous eigenvalue spectrum associated
to the FP equation, as discussed in Sec.\ \ref{tls:sec}, is vastly different from that
of the quantum double well system.

%%%%%%%%%%%%%%%%%%%%%%%%%%%%%%%%%%%%%%%%%%%%%%%%%%%%%%%%%%%%%%%%%%%%%%%%%
%                               BIBLIOGRAPHY
%%%%%%%%%%%%%%%%%%%%%%%%%%%%%%%%%%%%%%%%%%%%%%%%%%%%%%%%%%%%%%%%%%%%%%%%%


\begin{thebibliography}{99}
\bibitem{einstein:nota}
It is easy to see, by direct substitution in Eq.~(\protect\ref{FP:eqn}), that, 
at constant $T$, the Boltzmann distribution $e^{-V(x)/k_BT}$ is a stationary 
solution of the Fokker-Planck equation only if $\eta$ and $D$ satisfy 
the Einstein's relation, $\eta(T) D(T) = k_BT$.
\bibitem{timescale:nota}
If, for instance, $\epsilon_{res}(\tau)$ behaves in one scheme as a power-law, 
$1/\tau^{\alpha}$, and in another as a slowly decreasing logarithm, $1/\log{\tau}$, then 
the first scheme will be sooner or later more convenient independently of most details.
\bibitem{grid}
Typically, we restrict the variable $x$ in the interval $[-3,3]$,
beyond which the potential is too large and $P(x,t)$ is negligible,
and use a grid of up to 1000 points, checking for convergence of the
results.
\bibitem{fit}
This is generally speaking a difficult fit: The estimated value of $\Delta/B$ 
is between 0.14 and 0.22, depending on the inclusion of sub-leading corrections,
to be compared with the theoretical value of 0.22. 
Neglecting the log-corrections leads to a wrong estimate of the power-law 
exponent.
\bibitem{log_continuum:nota}
For the sake of precision, we should stress the fact that the logarithm
is, strictly speaking, born out of the continuum limit, which is indeed
valid only for $T>>\Delta=ka^2/2$. If the lattice constant $a$ is kept finite, 
there is a minimum non-zero value $\Delta_V$ for the splitting between the 
bottom of any two valleys, and, based on the two-level system results, we
could anticipate a final power-law behavior $\tau^{-\Delta_V/B}$. 
Nevertheless, the logarithmic behavior should apply in a whole temperature
window with a final crossover to the power-law behavior.
\bibitem{sign_V:nota}
If we consider that $\eta_t D_t=k_B T(t)$, we clearly see that, when annealing
the temperature, the overall sign of the factor in front of $V(x)$ in the
extra term of $V_{FP}$ is negative.
\end{thebibliography}
\end{document}